\documentclass[prb,superscriptaddress,showpacs,papersize=a4paper,twocolumn]{revtex4-1}%
\usepackage[pdftex]{graphicx}
\usepackage{bm}
\usepackage{etoolbox}
\usepackage[usenames,dvipsnames]{color}
\usepackage{mathtools}
\usepackage{amsmath}%
\usepackage{amsfonts}%
\usepackage{amssymb}
\DeclarePairedDelimiter\bra{\langle}{\rvert}
\DeclarePairedDelimiter\ket{\lvert}{\rangle}
\DeclarePairedDelimiterX\braket[2]{\langle}{\rangle}{#1 \delimsize\vert #2}

\DeclareMathOperator \sh {sh}
\DeclareMathOperator \ch {ch}
\newcommand{\bg}{ \begin{gather} }
\newcommand{\eg}{\end{gather}}
\newcommand{\be}{ \begin{equation} }
\newcommand{\ee}{\end{equation}}
\newcommand{\bea}{ \begin{eqnarray} }
\newcommand{\eea}{\end{eqnarray}}

\newcommand{\str}{\mathop{\rm Str}}

\begin{document}

\title{Multifractality of wave functions on a Cayley tree: From root to leaves }

\author{M.~Sonner}
\affiliation{Institut f{\"u}r Theorie der Kondensierten Materie, Karlsruhe Institute of Technology, 76128 Karlsruhe, Germany}

\author{K.\,S.~Tikhonov}
\affiliation{Condensed-matter Physics Laboratory, National Research University Higher
 School of Economics, 101000 Moscow, Russia}
\affiliation{Institut f{\"u}r Nanotechnologie, Karlsruhe Institute of Technology, 76021 Karlsruhe, Germany}
\affiliation{L.\,D.~Landau Institute for Theoretical Physics RAS, 119334 Moscow, Russia}

\author{A.\,D.~Mirlin}
\affiliation{Institut f{\"u}r Nanotechnologie, Karlsruhe Institute of Technology, 76021 Karlsruhe, Germany}
\affiliation{Institut f{\"u}r Theorie der Kondensierten Materie, Karlsruhe Institute of Technology, 76128 Karlsruhe, Germany}
\affiliation{L.\,D.~Landau Institute for Theoretical Physics RAS, 119334 Moscow, Russia}
\affiliation{Petersburg Nuclear Physics Institute,188300 St.\,Petersburg, Russia.}

\begin{abstract}
We explore the evolution of wave-function statistics on a finite Bethe lattice (Cayley tree) from the central site (``root'') to the boundary (``leaves'').
We show that the eigenfunction moments $P_q=N \left<|\psi|^{2q}(i)\right>$ exhibit a multifractal scaling $P_q\propto N^{-\tau_q}$ with the volume (number of sites) $N$ at $N\to\infty$. The multifractality spectrum $\tau_q$ depends on the strength of disorder and on the parameter $s$ characterizing the position of the observation point $i$ on the lattice. Specifically, $s= r/R$, where $r$ is the distance from the observation point to the root, and $R$ is the ``radius'' of the lattice. We demonstrate that the exponents  $\tau_q$  depend linearly on $s$ and determine the evolution of the spectrum with increasing disorder, from delocalized to the localized phase. Analytical results are obtained for the $n$-orbital model with $n \gg 1$ that can be mapped onto a supersymmetric $\sigma$ model. These results are supported by numerical simulations (exact diagonalization) of the conventional ($n=1$) Anderson tight-binding model. 
\end{abstract}
\maketitle

\section{Introduction}
\label{SectionIntro}

Anderson localization is one of the most fundamental quantum phenomena. In the conventional setting, it is formulated as a problem of non-interacting quantum particles moving in $d$ dimensions in a random potential. As was proven by Anderson \cite{anderson58}, for a sufficiently strong disorder the single-particle wave functions are spatially localized. Changing the disorder strength (or another control parameter) can drive the system from the localized to the delocalized phase. Such localization-delocalization transitions, known as Anderson transitions, exhibit a very rich physics\cite{evers08}. A  particularly remarkable feature of these transitions is the multifractality of critical wave functions. The fractal exponents characterizing the scaling of moments of wave functions are universal in the sense of universality of critical indices (i.e. for given spatial dimensionality $d$ and underlying symmetry and topology). The multifractality of eigenfunctions in a $d$-dimensional disordered system holds only at the transition point. 

Anderson localization has recently attracted a renewed interest in the context of transport and ergodicity in interacting disordered systems at non-zero temperature. The effect of interactions on localized excitations was addressed in an early paper, Ref.~\onlinecite{fleishman1980interactions},  where it was argued that the discretness of the spectrum may prohibit delocalization by inelastic processes if the interaction is sufficiently short-ranged. 
Two decades later it was demonstrated\cite{altshuler1997quasiparticle} that similar phenomena are also relevant for quantum dots. 
In this case, single-particle wave functions are extended over the whole system, and localization happens in the Fock space rather than in the real space. The authors of Ref.~\onlinecite{altshuler1997quasiparticle} suggested an approximate mapping between the problem of Fock-space localization in a quantum dot and a single-particle localization on a Bethe lattice. (See also subsequent works\cite{mirlin97,jacquod1997emergence,silvestrov97,weinmann97,georgeot1998integrability,silvestrov98,mejia-monasterio98,berkovits98,leyronas99,leyronas00,rivas02,gornyi2016many}.)
Later, these ideas were extended to explore the many-body-localization (MBL) in spatially extended systems with localized single-particle states and with short-range interaction\cite{gornyi2005interacting,basko2006metal}.
The conclusion of these works on the existence of a finite-temperature MBL transition was supported by numerous subsequent numerical and analytical studies, see, in particular, Refs.~\onlinecite{oganesyan07,monthus10,kjall14,gopalakrishnan14,luitz15,nandkishore15,karrasch15,imbrie16,imbrie16a,gornyi-adp16}. 
Recently, experimental realizations of 2D bosonic\cite{choi2016exploring} and 1D fermionic\cite{schreiber2015observation} systems showing MBL transition were implemented for cold atoms in disordered optical lattices. Signatures of MBL transition
in interacting systems were also observed  in InO films~\cite{ovadyahu1,ovadyahu2,ovadia2015evidence}. Further, the MBL was studied experimentally in arrays of coupled one-dimensional optical lattices~\cite{bordia15,lueschen16}.
The MBL transition takes place also in models with long-range interaction but the critical disorder in this case was predicted to show a non-trivial scaling with the system size\cite{Burin15,Gutman16}. Possible experimental realizations of such models include ultra-cold polar molecules or magnetic spin impurities in solid state\cite{Demler14,smith16}. 

The similarity between the interacting problems (with interaction matrix elements producing an effective hopping in the Fock space) and single-particle problems on tree-like lattices triggered a recent increase of interest to the latter models. In fact, the model of a quantum particle  hopping over a Cayley tree  with connectivity $K = m+1$ in a potential disorder provided a long ago \cite{abou1973selfconsistent} the first exact solution of a localization problem exhibiting an Anderson transition. The Hamiltonian of this model reads
\begin{equation}
\label{H}
\mathcal{H}=t\sum_{\left<i, j\right>}\left(c_i^+ c_j + c_j^+ c_i\right)+\sum_{i=1} \varepsilon_i c_i^+ c_i\,,
\end{equation}
where the sum is over the nearest-neighbour sites of the Cayley tree. The energies $\varepsilon_i$ are independent random variables sampled from a given distribution; the standard choice is a uniform distribution on $[-W/2,W/2]$. The analysis of Ref.~\onlinecite{abou1973selfconsistent} and its extensions in subsequent works \cite{efetov1985anderson,zirnbauer1986localization,zirnbauer1986anderson,efetov1987density,efetov1987anderson,verbaarschot1988graded,mirlin1991localization}
made it possible to establish the transition point and the corresponding critical behavior. These findings have been corroborated and supplemented by mathematically rigorous results \cite{klein98,aizenman11}.

The analysis of Refs.~\onlinecite{abou1973selfconsistent,efetov1985anderson,zirnbauer1986localization,zirnbauer1986anderson,efetov1987density,efetov1987anderson,verbaarschot1988graded,mirlin1991localization} was carried out in the limit of an infinite-size system. On the hand, many important observables---such as  statistical properties of eigenfunctions and of the energy spectrum---are defined only for a finite system. One natural finite-size
modification of the Cayley tree model is provided by the  sparse random matrix (SRM) ensemble (also known as Erd\"os-R\'enyi graphs in mathematical literature) studied analytically in Refs.~\onlinecite{mirlin1991universality,fyodorov1991localization,fyodorov1992novel}. A closely related model is that  of a random regular graph (RRG), which essentially represents a finite Cayley tree wrapped onto itself. The RRG and SRM ensembles are very similar tree-like models without boundary (and with loops of typical size $\sim \ln N$). 

 It was found in Refs.~\onlinecite{mirlin1991universality,fyodorov1991localization,fyodorov1992novel} that, in the limit of large number of sites $N$, the delocalized phase on the infinite cluster of the SRM model has ergodic nature in the sense of, first, the Wigner-Dyson level statistics, and, second, the $1/N$ scaling of the inverse participation ratio (IPR) $P_2=\sum_i|\psi_i|^4$ characterizing eigenfunction fluctuations on the infinite cluster. (Here $\psi_i$ is the amplitude of a wave function $\psi$  on site $i$.)
 More recently, the ergodicity of the delocalized phase in the RRG model was questioned in Refs.~\onlinecite{biroli12,deluca14}.  
These works motivated an intensive numerical research on properties of the delocalized phase in the RRG and SRM models \cite{tikhonov2016anderson,garcia-mata17,metz17,biroli2017delocalized}.  A detailed numerical investigation of level and eigenfunctions statistics on the delocalized side of the Anderson transition on RRG carried out in Ref.~\onlinecite{tikhonov2016anderson} supported the analytical prediction of Refs.~\onlinecite{mirlin1991universality,fyodorov1991localization,fyodorov1992novel}. More specifically, the numerical analysis of Ref.~\onlinecite{tikhonov2016anderson} reveals a crossover from relatively small ($N\ll N_c$) to large ($N\gg N_c$) systems, where  $N_c$ is the correlation volume. The values of $N_c$ obtained from numerical simulations are in good agreement with the analytical prediction $\ln N_c \sim (W_c-W)^{-1/2}$
implying an exponential divergence of the correlation volume at the transition point.  For $N\ll N_c$ the system exhibits  a flow towards the Anderson-transition fixed point which has on RRG properties similar to the localized phase. Only when the system volume $N$ exceeds $N_c$, the direction of flow is reverted and the system approaches its $N\to\infty$ ergodic behavior. This  non-monotonous behavior, along with exponentially large values of the correlation volume $N_c$, makes the finite-size analysis very non-trivial.
The key conclusions of Ref.~\onlinecite{tikhonov2016anderson} have been corroborated by subsequent numerical studies of the RRG and SRM models \cite{garcia-mata17,metz17}. 

Thus, properties of the delocalized phase on RRG---in particular, the ergodicity manifesting itself on scales $N\gg N_c$---are now largely understood, both analytically and numerically. 
There is, however,  another natural way to define a finite-size model related to the Bethe lattice. Specifically, one can simply keep only sites with a certain distance $R$ from a given site (known as ``root''), which yields a finite Cayley tree. Until recently,
properties of delocalized eigenfunctions on a finite Cayley tree have remained largely unexplored.  We recall that the original analysis of a hopping problem on a Cayley tree was performed on an infinite lattice. This formulation is appropriate for determination of the position of the transition point and for investigation of properties of localized wave functions as well as of finite-frequency correlation functions in the delocalized phase. When a ``linear size'' ($\sim \ln N$)  of the lattice is much larger than $L_\omega$ (a characteristic displacement of a particle in a time $\sim 1/\omega$ for a given frequency $\omega$), the boundary conditions should not play a role and the system can be considered as infinite. 
On the other hand, for the case of such observables as 
the statistics  of eigenfunctions and energy levels on the delocalized side of the transition, the situation is more intricate.  Indeed, at variance with finite-frequency correlation functions, the mere definition of such observables requires a consideration of a finite system. Since most sites of a Cayley tree are located at the boundary, one can expect that the presence of boundary may crucially affect the wave-function and level statistics in the delocalized phase.
Indications of a peculiar character of eigenstates on a Cayley tree were provided by Monthus and Garel \cite{monthus2008anderson,monthus2011anderson} who studied numerically the statistics of transmisson amplitudes on a Cayley tree in a scattering geometry. Recently, two of the present authors studied, both analytically and numerically, the statistics of eigenfunctions in the root of a Cayley tree\cite{tikhonov2016fractality}. Our results proved that---in line with above expectations of a role of boundary conditions---the statistics on the Cayley tree is qualitatively different from that on RRG. Specifically, we have shown that the eigenfunction amplitudes at the root of the tree are distributed fractally in the most of the extended phase, which should be contrasted to the ergodicity of the extended phase in the RRG model.

Interestingly, it is known on the mathematical level of rigor\cite{aizenman2006canopy} that that random Schr\"odinger operators on so-called canopy graphs have pure-point spectrum for any strength of disorder, at least for some models of disorder distribution.
A canopy graph is an infinite tree that represents a $R\to \infty$ limit of a sequence of Cayley trees of ``radius'' $R$ from the perspective of a boundary site.
 This suggests localization of eigenstates near the boundary of a Cayley tree.  It is then natural to ask whether this localization is compatible with eigenfunction statistics at the root characteristic for delocalized (ergodic or fractal) regime, as found in Ref.~\onlinecite{tikhonov2016fractality}. 

The goal of the present work is to explore the wavefunction statistics at an arbitrary location on a finite Cayley tree, from the root of the tree to the ``leaves'' (i.e., sites located on the boundary). We will show that, very generally, this statistics is of multifractal character and will determine the corresponding spectrum of multifractal exponents. Our investigation combines analytical and numerical methods. Specifically, we will first use the supersymmetric $\sigma$-model approach to study analytically the $n$-orbital model with $n\gg 1$ (Sections \ref{s2} and \ref{SectionAnalytical}).  Next, we will analyze  numerically the conventional $n=1$ Anderson model in  Sec.~\ref{numerics}. We will show that the results for both models agree  with each  other very well, i.e., most of the essential features of the $n=1$ Anderson model are fully captured by the $n \gg 1$ limit. Our central result is the evolution of the multifractal spectrum with the position on a tree and with the strength of disorder. Interestingly, we find that some multifractal features hold even in the localized phase. Further, while our results imply the localized character of states at the boundary (which is in agreement with Ref.~\onlinecite{aizenman2006canopy}), we will show that certain moments of wave functions retain their multifractal behavior in this case as well.

\section{Recurrence relations for n-orbital model}
\label{s2}

\subsection{Supersymmetry formalism}

In order to simplify the analytical treatment of the problem, we will consider the  $n$-orbital generalization of the problem defined by Hamiltonian (\ref{H}), with $n\gg 1$. The Hamiltonian of such a ``granular'' system reads:
\begin{eqnarray}
\mathcal{H} &=&
\sum_{\left<i, j\right>}\sum_{p,q=1}^n \left(t_{ij}^{pq} c_{ip}^+ c_{jq}  + {\rm h.c.} \right) 
\nonumber \\ 
&+&
\sum_{i}\sum_{p,q=1}^n (\varepsilon_i^{pq} c_{ip}^+ c_{iq} + {\rm h.c.}). \label{HG}
\end{eqnarray}
Here $i,j$ label sites (``granules''), and $p,q$ states (``orbitals'') belonging to each of them. For large $n$, the $n$-orbital problem can be mapped onto a supersymmetric $\sigma$-model \cite{efetov1985anderson,zirnbauer1986localization,zirnbauer1986anderson,efetov1987density,efetov1987anderson,verbaarschot1988graded}. 
The derivation of the $\sigma$-model is simplified if one assumes that $t_{ij}^{pq}$ and  $\varepsilon_i^{pq}$ are Gaussian distributed random variables; the mapping applies, however, under much more general conditions. In physical terms, the underlying condition is the ergodicity on the scale of a single granule. Each of the granules $i$ is then described separately by a zero-dimensional $\sigma$-model. Hopping between the granules [the first term in the Hamiltonian (\ref{HG})] couples these zero-dimensional models into a sigma-model on the lattice formed by sites $i$ [see the action (\ref{action}) below]. The mapping works for any geometry of the lattice; in the case of our interest in the present paper, this is a finite Cayley tree. 

For  the $n=1$ Anderson model on an infinite Bethe lattice the supersymmetry solution was achieved in Ref.~\onlinecite{mirlin1991localization}.
While the $n=1$ model 
and its $n\gg 1$ generalization ($\sigma$-model) turn out to exhibit the same gross features, analytical calculations are somewhat simpler within the $\sigma$-model. For this reason, we find it instructive to carry out the analytical investigation within the $n\gg 1$ model (i.e., the $\sigma$-model).

We will assume free boundary conditions (corresponding to an isolated system) and study statistics of wavefunction amplitudes $u_i=|\psi_i|^2$ at a site $i$ located on the distance $r$ from the root of the tree. As the sites of a finite tree are clearly not equivalent, the distribution function of $u_i$ depends on $r$.   As we show below, the proper scaling variable characterizing the position of the point on the Cayley tree in the limit $N\to\infty$ is  $s=r/R$, where $R$ is the ``radius'' of the tree. Specifically, we will show that $s$ controls (for a given strength of disorder) the spectrum of multifractal exponents. According to the definition, we have $0 \le s \le 1$,  with $s=0$ corresponding to the root and $s=1$ to leaves (i.e., the boundary). In the case of $s=0$, this problem was studied in the previous paper by two of us\cite{tikhonov2016fractality} where it was demonstrated that the distribution function of amplitudes $|\psi_i|^2$ at the root exhibits a transition from ergodic to fractal behaviour as a function of disorder inside the delocalized phase. In the present work, we will study the eigenfunction statistics for all $s$ and for all disorder strengths. We will see that the results of Ref.~\onlinecite{tikhonov2016fractality}, as well the localization at the boundary found in Ref.~\onlinecite{aizenman2006canopy}, are limiting cases of a general picture of multifractality on a Cayley tree. 

A powerful analytical approach to the eigenfunction statistics  of a non-interacting disordered system is the supersymmetry method, see Ref.~\onlinecite{mirlin2000statistics} for detail. The starting point is the expression for moments of wavefunction amplitudes  at a given node $j$ and energy $\varepsilon$:
\be
\label{def}
\left<|u_j|^q\right> = \frac{i^{q-2}}{2\pi\nu N} \lim_{\eta\to 0} (2\eta)^{q-1}\left<G_R^{q-1}(j)G_A(j)\right> \,,
\ee
where $G_{R(A)}(j)$ are retarded and advanced Green functions at coinciding points,
\be
G_{R(A)}(j)=\bra{j}\left(\varepsilon-\hat H\pm i\eta\right)^{-1}\ket{j} \,,
\label{Green}
\ee
$\hat H$ is the Hamiltonian, and $\nu$ is the density of states at energy $\varepsilon$. Next steps include expressing the Green functions in terms of intergrals over a supervector field and averaging over disorder.
In the case of a model with $n\gg 1$ orbitals per lattice site, the resulting supersymmetric theory acquires a form of the  $\sigma$-model\cite{efetov1985anderson} with the action
\be
\label{action}
S[Q]= - J\sum_{\left<i, j\right>}\str\left[Q(i)-Q(j)\right]^2 + \frac{\pi\eta}{2\delta_0}\sum_i \str\left[\Lambda Q(i)\right].
\ee
Here $Q(j)$ are a $8 \times 8$ supermatrices associated with lattice sites $j$ and satisfying the condition $Q(j)^2=1$, the symbol $\str$ denotes the supertrace (i.e., the trace of the boson-boson block minus trace of the fermion-fermion block), 
$\Lambda$ is a diagonal matrix with entries equal to 1 and -1 for the retarded and advanced subspaces, respectively,
$\delta_0 = \nu^{-1} $ is the mean level spacing on a site, and $J=\left(t/\delta_0\right)^2$ is the dimensionless coupling constant. Here $t$ is the characteristic amplitude of the hopping, $t^2 = \langle |t_{ij}^{pq}|^2\rangle$. If all amplitudes are real, the Hamiltonian (\ref{HG}) belongs to the orthogonal (AI) symmetry class, which determines the associated symmetry of the  $\sigma$ model. If the time-reversal symmetry is broken (e.g., the hopping amplitudes $t_{ij}^{pq}$ are made complex with random phases), the symmetry class changes from orthogonal to unitary (A). While all essential features of the wave function statistics discussed in this paper are the same in both cases, the unitary-symmetry case is somewhat simpler technically. Thus, for the sake of transparency of exposition, we will focus on the model of the unitary symmetry class below. In this case, $Q(i)$ in Eq.~(\ref{action}) are $4\times 4$ supermatrices, and the action  (\ref{action}) acquires an additional overall factor of two.

Expressing the average product of Green functions in Eq.~(\ref{def}) in terms of a sigma-model correlation function, one brings  the expression for the moments of wave functions to the following form\cite{mirlin2000statistics}:
\bea
\label{srep}
\left<|u_i|^q\right> &=& -\frac{q}{2N}\lim_{\eta\to 0}(2\pi\eta/\delta_0)^{q-1} \nonumber \\
& \times& \int DQ \:Q^{q-1}_{11,bb}(i)Q_{22,bb}(i)\:e^{-S[Q]}.
\eea
Here the preexponential factor is a product of elements of the matrix $Q(i)$ at the point $i$ of the Cayley tree where the eigenfunction statistics is studied. The first two indices of $Q$ correspond to the advanced-retarded and the last two to the boson-fermion decomposition.

%%%%%%%%%%%%%%%%%%%%%%%%%%%%%%%%
\begin{figure}
\centering
\includegraphics[width=0.5\textwidth]{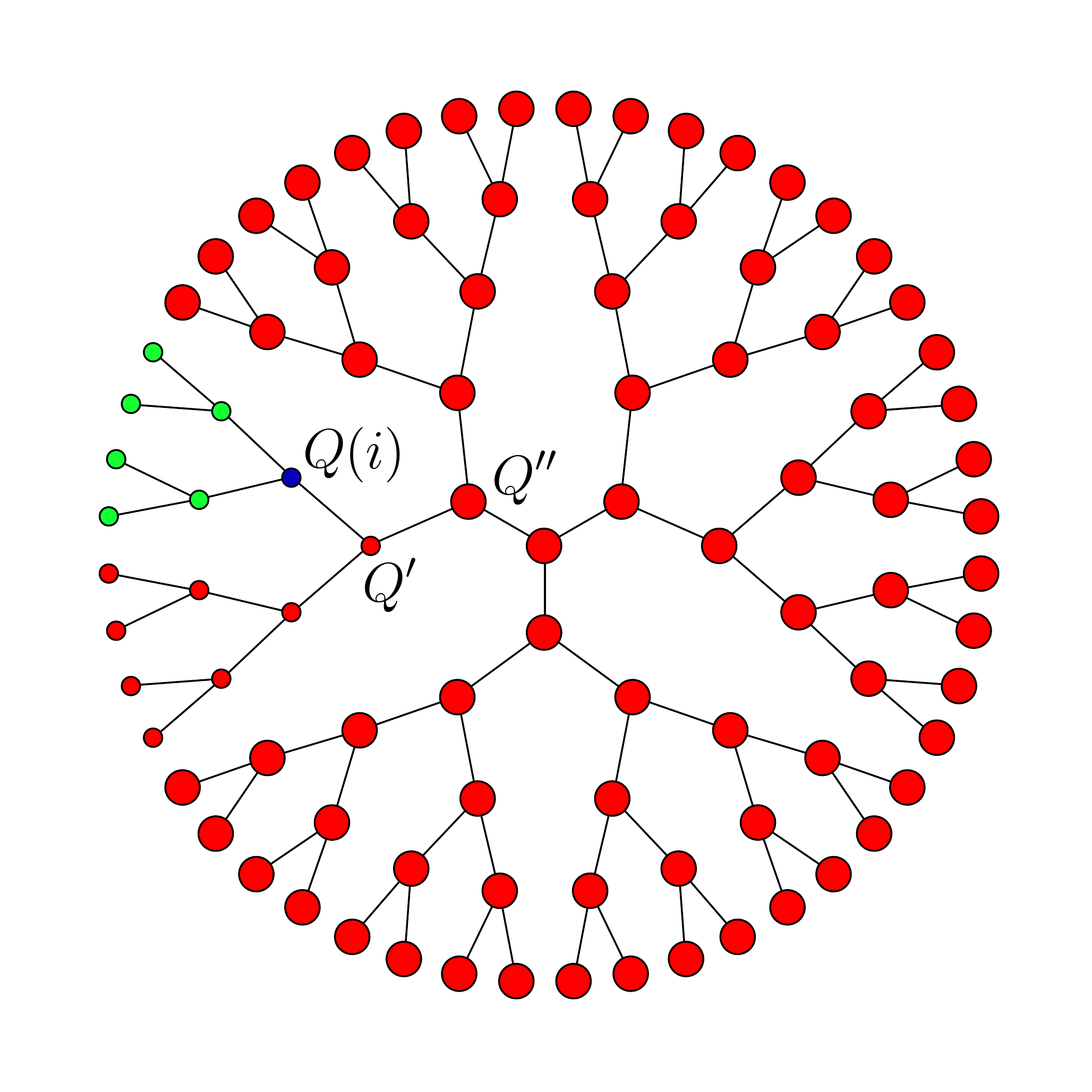}
\label{ct1}
\caption{Cayley tree: Evaluation of the integral $Y(Q(i))$ over all nodes except for  $Q(i)$ (blue node) for $m=2,\;r=3,\;R=5$. Integration over the green nodes gives $\Psi^{m}_{R-r}(Q(i))$, integration over all red nodes gives $\Xi_{R,r}(Q(i))$. This yields Eq.~(\ref{Yr}). Partioning of red nodes into large and small illustrates the recurrence equation for $\Xi$, Eq.~(\ref{PiRecursion}): the integral over large red nodes yields $\Xi_{R,r-1}(Q')$.}
\label{ct}
\end{figure}
%%%%%%%%%%%%%%%%%%%%%%%%%%%%%%%%%%%%

For a generic lattice, an analytic evaluation of the functional integral in Eq.~(\ref{srep}) is a formidable task. The tree structure of the lattice greatly simplifies it, making it possible to express the result in terms of solutions of certain recursive equations, as illustrated in Fig.~\ref{ct}. The site $i$ where the statistics is evaluated is blue colored in the figure. The statistics of wave functions at this site is determined, according to Eq.~(\ref{srep}), by the result of integration over all other nodes. This integral, which we denote $Y(Q(i))$,   is factorized:
\be
Y_r(Q(i))=\Psi^{m}_{R-r}(Q(i))\Xi_{R,r}(Q(i)).
\label{Yr}
\ee 
Here $\Psi^{m}_{R-r}(Q)$ is a result of integration over $m$ branches (green-colored in the figure) which ``grow'' from the site $i$ towards the boundary. Each such branch yield a factor $\Psi_{R-r}(Q)$, with the subscript $R-r$ indicating the distance between the site $i$ (which plays a role of the root for these branches) to the boundary.  The second factor $\Xi_{R,r}(Q)$ in Eq.~(\ref{Yr}) is a result of integration over the remaining (red-coloured) nodes of the lattice.  It is important that both the functions $\Psi_{R-r}(Q)$ and $\Xi_{R,r}(Q)$ can be found recursively. The first of them satisfies 
the following recurrence relation:
\be
\label{PsiRecursion}
\Psi_{r+1}(Q)=\int DQ' e^{-\str\left[-2J(Q-Q')^2+\frac{\pi\eta}{\delta_0}\Lambda Q'\right]}\Psi_{r}^m(Q'),
\ee
with the initial condition $\Psi_0(Q) = 1$. 

The recursive relation for $\Xi(Q)$ is slightly more involved. In order to derive it, we consider again a site $i$ with a matrix $Q(i)$ and denote by $Q'$ the matrix associated with its parent site, see Fig.~\ref{ct}. Further, we 
consider a subtree growing from $Q'$ towards the boundary and cut off a branch containing $Q(i)$. This defines a set of ``small'' red nodes shown on the Fig.~\ref{ct} by small red circles. The remaining red nodes are termed ``large'' ones and are shown by large red circles in Fig.~\ref{ct}. The integral over the large nodes is exactly $\Xi_{R,r-1}(Q')$. The integral over small red nodes except for the one directly coupled to site $i$ (with the matrix $Q'$) yields $\Psi^{m-1}_{R-r+1}(Q')$.
Therefore, we find the recurrence relation
\bea
\label{PiRecursion}
\Xi_{R,r}(Q)&=&\int DQ' e^{-\str\left[-2J(Q-Q')^2+\frac{\pi\eta}{\delta_0}\Lambda Q'\right]} \nonumber \\
& \times& \Xi_{R,r-1}(Q')\Psi^{m-1}_{R-r+1}(Q') \,,
\eea
with the boundary condition $\Xi_{R,0}(Q)=\Psi_R(Q)$.

As is clear from Eq.~(\ref{srep}), the distribution function of the wave function intensity $u_i$ at the given node $i$ is fully determined by the
asymptotic form of the function $Y_r(Q(i))$, resulting from integrating out all degrees of freedom on the tree except for the matrix $Q(i)$ at the node under interest. 

The set of equations which we have just derived allows one to evaluated the functional integral in Eq.~(\ref{srep})  at arbitrary value of the level broadening $\eta$. Being interested in properties of an isolated system, we now turn to a discussion of the limit of $\eta\to 0$ that determines the eigenfunction statistics according to Eq.~(\ref{srep}). 
 
\subsection{From supersymmetry to real numbers}

To proceed further, we note that, in view of the symmetry of the $\sigma$-model action, the functions $\Psi_r(Q)$, $\Xi_{R,r}(Q)$, and $Y_r(Q)$  in the unitary symmetry case depend only on two variables $1\leq\lambda_1<\infty$ and $-1\leq\lambda_2\leq 1$, which are the eigenvalues of the retarded-retarded block of the matrix $Q$, see Ref.~\onlinecite{mirlin2000statistics}. The variables $\lambda_1$ and $\lambda_2$ correspond to the non-compact (hyperbolic) and compact (spherical) sectors of the $\sigma$-model coset space.  As we are interested in the limit of $\eta \to 0$ at fixed $N$ (and hence at fixed $n$ and $r$), we can further simplify the equations (\ref{PiRecursion}) and (\ref{PsiRecursion}). Specifically, in this limit only the dependence on $\lambda_1$ persists:
\be
\Psi_r(Q)\equiv \Psi_r(\lambda_1,\lambda_2) \to \Psi_r^{(a)}(2\pi\eta\lambda_1/\delta_0),
\label{asymptotic}
\ee
where the superscript $(a)$ indicates that we are dealing with the asymptotic, small-$\eta$ form of the function $\Psi_r$. A fully analogous reduction holds for the functions $\Xi_{R,r}(Q)$ and $Y_r(Q)$.  The function 
\be 
Y^{(a)}_r(u)  =   \left[\Psi^{(a)}_{R-r}(u)\right]^m\Xi^{(a)}_{R,r}(u)
\label{Yar}
\ee
determines, in view of Eq.~(\ref{srep}) the eigenfunction moments and thus the distribution function ${\cal P}(u_i)$. Specifically, one finds\cite{mirlin2000statistics}
\be
\mathcal{P}(u_i)=N^{-1}\partial_u^2 Y^{(a)}_{r}(u_i).
\label{Pu}
\ee

In the $\eta\to 0$ limit, in which the functions $\Psi_r$ and $\Xi_{R,r}$ depend on a single scalar variable [see Eq.~(\ref{asymptotic})], the recurrence relations (\ref{PsiRecursion}) and  (\ref{PiRecursion}) can be substantially simplified. It is convenient to introduce $t = \ln(2\pi\eta\lambda_1/\delta_0)$ and to perform the corresponding change of variable
\be
\label{Phi}
\Psi_r^{(a)}(e^t) = \psi_r(t)\,, \qquad \Xi_{R,r}^{(a)}(e^t) = \xi_{R,r}(t) \,.
\ee
Equation (\ref{PsiRecursion}) reduces then to the following asymptotic recurrence relation for $\psi_r(t)$:
\be
\label{PsiRecursionSimple}
\psi_{r+1}(t)=\int L(t-t')e^{-e^{t'}}\psi^m_r(t')dt',
\ee
where the kernel $L(t)$ is given by
\be
L(t)=\frac{2g\ch g+(2g\ch t-1)\sh g}{2\sqrt{2\pi g}}e^{t/2-g\ch t},
\label{Lt}
\ee
with $g=8J$. The starting point of this recursion is 
\be 
\psi_0(t) = 1. 
\ee
In a similar way, Eq.~(\ref{PiRecursion}) takes the asymptotic form of a recurrence relation for $\xi(t)$:
\be
\label{PiRecursionSimple} 
\xi_{R,r}(t)=\int L(t-t')e^{-e^{t'}} \xi_{R,r-1}(t') \psi^{m-1}_{R-r+1}(t')dt'\,,
\ee
with the initial condition 
\be 
\label{xi-initial}
\xi_{R,0}(t) = \psi_{R}(t).
\ee
It is worth reminding the reader that within the $\sigma$-model formalism (corresponding to the $n$-orbital model with $n \gg 1$) all the information about the degree of disorder (and thus of localization) is encoded in the coupling constant $g$. 

The solution of the recurrence relations (\ref{PsiRecursionSimple}) and (\ref{PiRecursionSimple}) yield, in view of Eqs.~(\ref{Yar}) and (\ref{Phi}), the function $Y^{(a)}_r(u)$.  This function determines,  by virtue of Eq.~(\ref{Pu}), the distribution of the wave function amplitudes and thus the corresponding moments. However, the analysis of the resulting behavior of functions $\psi_r(t)$ and $\xi_{R,r}(t)$ as well as of the wave function statistics is far from trivial in view of non-linearity of the recurrence.

\section{From recurrence relations to wave function multifractality}
\label{SectionAnalytical}

In this Section, we will perform a detailed analysis of the solution of recursive relations (\ref{PsiRecursionSimple}) and (\ref{PiRecursionSimple}).
Further, we will explore implications of the properties of this solution for the statistics of wave functions. This analysis will be carried out in the full range of the $\sigma$-model coupling constant $g$. 

An important role for the analysis  is played by the spectrum of the linear integral operator with the kernel $L(t-t')$, where $L(t)$ is given by Eq.~(\ref{Lt}). Eigenfunctions of this operator are $e^{\beta t}$, and the corresponding eigenvalues $\epsilon_{\beta}$ read\cite{efetov1985anderson,zirnbauer1986localization,zirnbauer1986anderson}
\be
\label{specUnitary}
\epsilon_{\beta}=\frac{2g K_{\beta+1/2}(g)\sh g + 2K_{\beta-1/2}(g)(g\ch g-\beta\sh g)}{\sqrt{2\pi g}},
\ee
where $K_\nu(g)$ is the modified Bessel function of the second kind. The function $\epsilon_\beta$ satisfies $\epsilon_1 = \epsilon_0 = 1$ and $\epsilon_\beta = \epsilon_{1-\beta}$,  and has a minimum at $\beta=1/2$. The relevant values of $\beta$ belong to the interval $1/2 \le \beta \le 1$ where the function $\epsilon_\beta$ monotonously increases.

\subsection{%Analysis of the recurrence relation for 
$\psi_r(t)$ and wave function moments at the root}
\label{DetPsi}

We begin by  discussing  properties of the solution of Eq.~(\ref{PsiRecursionSimple}). A detailed analysis of this problem was carried out in the recent paper\cite{tikhonov2016fractality} by two of us. Here we briefly recall the main results that will be used in the present work. 

The function  $\psi_r(t)$ has a form of the kink with the asymptotic values $\psi_r \to 1$ and  $\psi_r \to 0$ on the left and right side of the kink, respectively. More specifically, there are three asymptotic regimes of the behavior of the function $\psi_r(t)$ at large $r$.
At negative $t$ with large enough $|t|$ the recursive equation can be linearized with respect to $1- \psi_r(t)$. The $t$ dependence of the solution in this range is determined by the factor   $e^{-e^{t'}} \simeq 1 - e^{t'}$   in the integrand of Eq.~(\ref{PsiRecursionSimple}). 
This yields, in view of $\epsilon_1=1$, 
\be
\psi_r(t) \simeq 1-e^{t+r\ln m}, \ \ \ t\lesssim t_-(r).
\label{small-t}
\ee
On the right side of the kink,  $t\gtrsim t_+(r)$, the form of the symmetry-breaking term $e^{-e^{t'}}$ guarantees a very fast decay of the function $\psi_r$, so that it can be safely replaced by zero for the purposes of our analysis:
\be
\psi_r(t) \simeq 0, \ \ \ t\gtrsim t_+(r).
\label{large-t}
\ee
Finally, in the intermediate region  one gets
\be
\psi_r(t) \simeq 1-ce^{\beta_*(t+r\alpha_*\ln m)},  \ \ \ t_-(r)\lesssim t \lesssim t_+(r),
\label{intermediate-t}
\ee
  where $c$ is a numerical constant, $\beta_*$ satisfies $1/2 \le \beta_* \le 1$, and 
\be
\label{alpha}
\alpha_*=\frac{\ln (m\epsilon_{\beta_*})}{\beta_*\ln m}.
\ee 
While, in analogy with Eq.~(\ref{small-t}), the asymptotics (\ref{intermediate-t}) can be found as a solution of the equation equation (\ref{PiRecursionSimple}) linearized around $\psi_r=1$, the exponent of $t$ dependence of $1- \psi_r(t)$ is modified in comparison with its bare value, $e^t \to e^{\beta_*t}$, due to nonlinear effects. 
The borders $t_-(r)$ and $t_+(r)$ between the asymptotic regimes can be found by matching the respective asymptotics and read as follows:
\be
t_-(r)=-\frac{1-\alpha_*\beta_*}{1-\beta_*}r\ln m, \qquad t_+(r)=-\alpha_* r\ln m.
\label{t-borders}
\ee
Clearly, $t_+(r)$ determines the position of the kink as defined by the point where the function $\psi_r$ crosses the value 1/2 (or any other fixed value between 0 and 1). According to Eq.~(\ref{intermediate-t}), the velocity with which the kink moves to the left is 
\be
v_{\beta_*}  = \alpha_* \ln m \equiv \frac{\ln (m\epsilon_{\beta_*})}{\beta_*}\,.
\label{velocity}
\ee

Let us now explain how the exponent $\beta_*$ is selected. The mechanism for this is different in the localized ($g < g_c$) and delocalized ($g>g_c$) phases, where $g_c$ is the critcial value of the coupling constant corresponding to the Anderson transition. In the localized phase, the kink does not move in the limit of large $r$, which implies that $\alpha_*=0$ and thus
\be
\label{beta-loc}
m\epsilon_{\beta_*}=1 \,, \qquad g < g_c.
\ee
The delocalized phase is in turn subdivided in two parts \cite{tikhonov2016fractality}: (i) $g_c < g < g_e$ in which the wave functions at the root show fractal properties, and (ii) $g > g_e$ where wave functions at the root are ergodic. In the first of them (``fractal at the root''), the exponent $\beta_*$ is selected by the condition of the minimal velocity,
\be
\label{beta-frac}
\left. \frac{\partial}{\partial\beta} \frac{\ln (m\epsilon_{\beta})}{\beta} \right|_{\beta= \beta_*} = 0 \,, \qquad g_c < g < g_e\,.
\ee
In the second part of the delocalized phase (``ergodic at the root''), the exponent $\beta_*$ takes its maximal possible value,
\be
\label{beta-erg}
\beta_* = 1 \,, \qquad g > g_e \,.
\ee
Note that in the latter regime $\alpha_*=1$  as follows from Eq.~(\ref{alpha}), while the ratio $(1-\beta_*)/(1-\alpha_*\beta_*)$ in Eq.~(\ref{t-borders}) is understood as equal to unity (which is its limit at $\beta_* \to 1$).

\begin{figure}
\centering
\includegraphics[width=.5\textwidth]{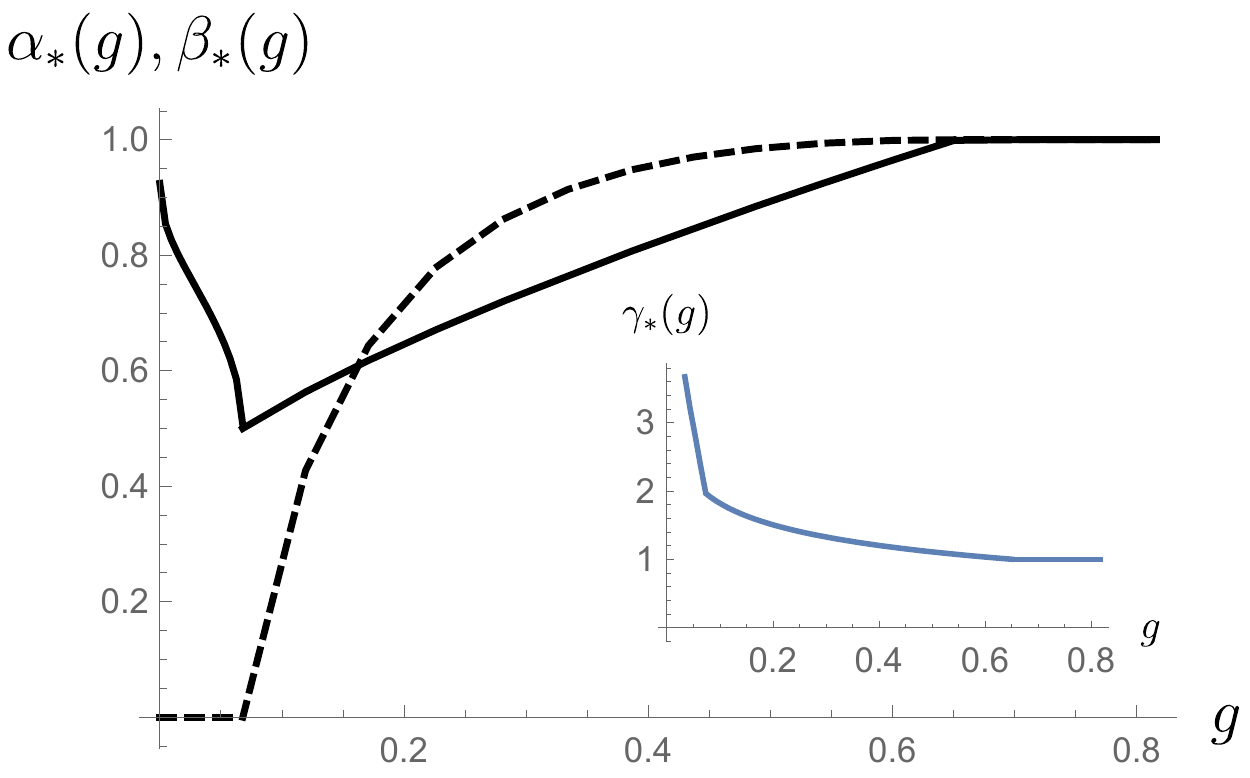}
\caption{Exponents $\beta_*(g)$ (solid) and $\alpha_*(g)$ (dashed) characterizing the drifting kink in $\psi_r(t)$. The branching number of the Cayley tree is chosen to be $m=2$. The Anderson-transition point $g_c \simeq 0.068$ separates the localized phase ($g<g_c$) with $\alpha_* = 0$, i.e. zero drift velocity, and the delocalized phase ($g>g_c$) where the drift velocity of the kink is non-zero, i.e., $\alpha_* > 0$. The point $g_e \simeq 0.65$ subdivides the delocalized phase into parts with ergodic ($g>g_e$, with $\alpha_*=1$) and non-ergodic ($g_c<g<g_e$, with $0<\alpha_*<1$) wave functions at the root. Inset: $\gamma_*$ as a function of the coupling constant $g$.
}
\label{figlambda}
\end{figure}

In Fig. \ref{figlambda} we show $\alpha_*$ and $\beta_*$ as functions of the coupling constant $g$ for a Cayley tree with branching number  $m=2$.   The Anderson transition point is then $g_c = 0.068$, while the border between the parts of the delocalized phase ergodic and non-ergodic at the root is $g_e = 0.65$. The non-analyticity of dependences $\alpha_*(g)$ and $\beta_*(g)$ at these points---which will induce also non-analytic behavior of the multifracality spectrum---is clearly seen.  As we discuss below, the analysis of the function $\xi_{R,r}(t)$ reveals the existence of another singular point located between $g_c$ and $g_e$ which manifests itself in the statistics of wave functions away from the root.

%%%%%%%%%%%%%%%%%%%%%%%
%%%%%%%%%%%%%%%%%%%%%%%%

%%%%%%%%%%%%%%%%%%%%%%%
\begin{figure*}
\centering
\includegraphics[width=\textwidth]{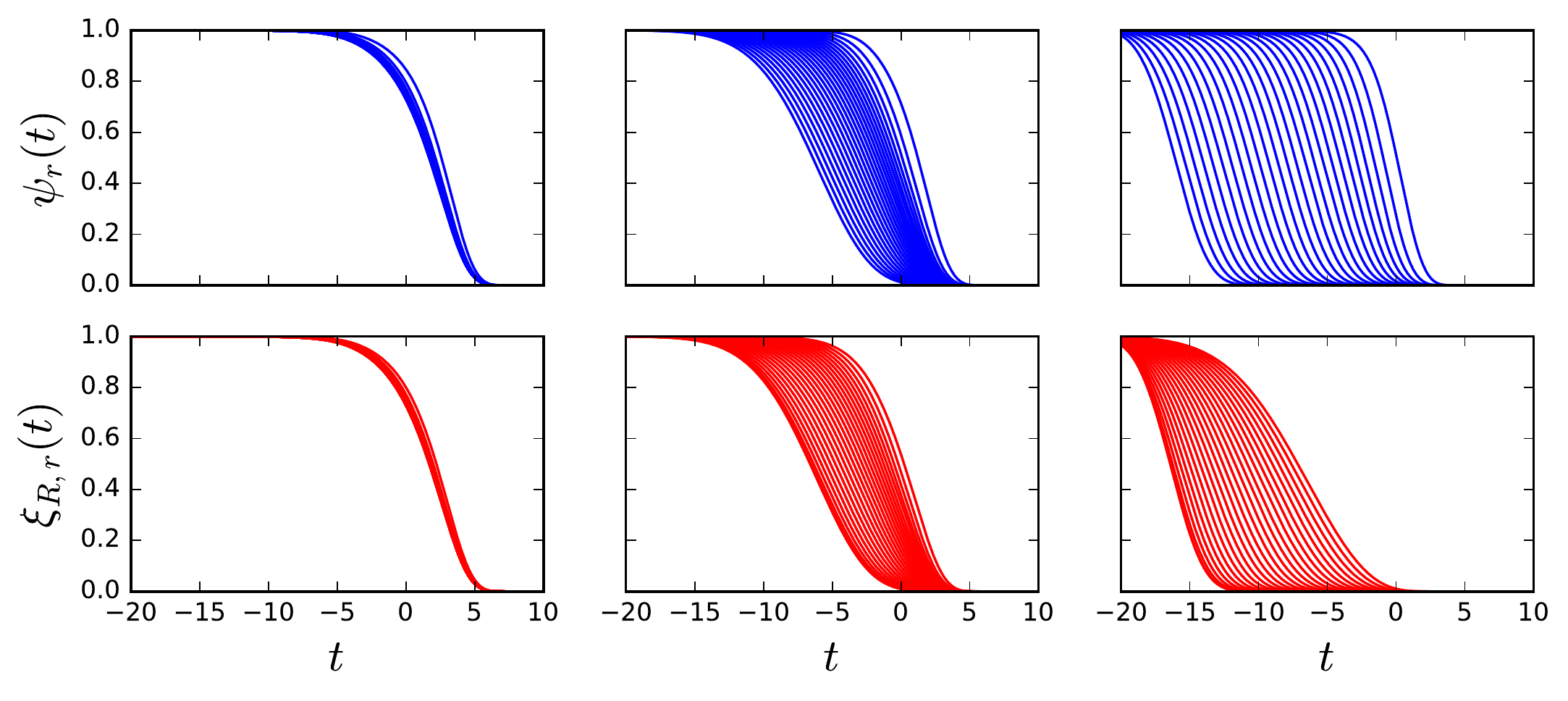}
\caption{Solution of the recursive relations (\ref{PsiRecursionSimple}) and (\ref{PiRecursionSimple}) for $\psi_r(t)$ (blue, upper row) and $\xi_{R,r}(t)$ (red, lower row) for the Cayley tree with a connectivity $m=2$ and radius $R=25$. {\it Left:}  The coupling constant is $g=0.05$ and satisfies $g<g_c$, the system is in the localized phase, the kink does not drift. {\it Middle:} The coupling constant is $g=0.2$ and satisfies $g_c < g < g_1 < g_e$; the system is in the delocalized phase, non-ergodic at the root; the kink in $\psi_r(t)$ moves with a velocity corresponding to $0 < \alpha_* < 1$; the kink in $\xi_{R,r}(t)$ moves back with the same velocity. {\it Right:} The coupling constant is $g=1$ and satisfies $g > g_e$; the system is in the delocalized phase, ergodic at the root; the kink in $\psi_r(t)$ moves with a maximum velocity corresponding to $\alpha_* = 1$; the kink in $\xi_{R,r}(t)$ moves back more slowly and gets strongly deformed.  
}
\label{fig-kinks}
\end{figure*}
%%%%%%%%%%%%%%%%%%%%%%%%

These results are illustrated in Fig. \ref{fig-kinks} where solutions of the recursion relation (\ref{PsiRecursionSimple}) for $\psi_r(t)$ are shown (blue curves) for the Cayley tree with a connectivity $m=2$ and radius $R=25$. To solve numerically the recursion relation, we  discretized  the integral recursions 
with mesh points chosen to be equally spaced in $t$, with 1000 points covering the range $t=[-40,40]$. Three panels  in the upper row of this figure show $\psi_r(t)$ for values of the coupling constant $g$ belonging to three different phases. In the left panel  the coupling is $g=0.05$ and the system is thus in the localized phase, $g < g_c$. Indeed, we see that that the kink does not drift in the large-$r$ limit, as expected. The middle panel  corresponds to the coupling $g=0.2$, so that the system is in the non-ergodic-at-root part of the delocalized phase, $g_c < g < g_e$. In this case the kink does drift, and the velocity corresponds to a certain $\alpha_*$ satisfying $0 < \alpha_* < 1$. Finally, in the right panel the coupling is $g=1$ and thus satisfies $g>g_e$, so that the system belongs to the ergodic-at-root part of the delocalized phase. The blue kink representing $\psi_r(t)$ moves with the maximum possible velocity, $\alpha_*=1$.

In Fig. \ref{fig:schematic_kinks}, black lines show schematically $\ln (1- \psi_R(t))$ and $\ln (1- \psi_{R-r}(t))$. In such a plot each of the regimes (\ref{small-t}), (\ref{large-t}), and (\ref{intermediate-t}) is represented by a straight line; the corresponding slopes are 0 for $t > t_+$, unity for $t<t_-$, and $\beta_*$ in the intermediate regime.

Using these results for $\psi_R$, one can evaluate, by using  Eq. (\ref{Pu}), the distribution of the wavefunction amplitudes $\mathcal{P}(u)$ and thus the wave function moments 
\be
\label{Pq}
P_q= N \left<|u_i|^q\right> = N \int du \, u^q \mathcal{P}(u). 
\ee
The result is that the important part of the distribution $\mathcal{P}(u)$ is of the power-law form, $\mathcal{P}(u) \propto u^{\beta_*-2}$ originating from Eq.~(\ref{intermediate-t}). As a consequence,  all moments with $q>q_*$, where
\be
\label{qstar}
q_*=1-\beta_*,
\ee
 are determined by the upper border of this power-law behavior, $u \sim N^{-\alpha_*}$ (originating from $t\sim t_+(R)$), while all moments with $q<q_*$ are determined by the lower border, $u\sim N^{-(1-\alpha_* \beta_*)/(1-\beta_*)}$  (originating from $t\sim t_-(R)$).  This yields the following scaling of the moments:
 \be 
 P_q\propto N^{-\tau_q},
 \label{Pq-scaling}
 \ee 
 with exponents $\tau_q$ given by\cite{tikhonov2016fractality}
\be
\label{qres}
\tau_q=\begin{cases}
\begin{array}{ll}
\displaystyle q\gamma_*-1, & \qquad q<q_*\:; \\[0.3cm]
\alpha_*(q-1), & \qquad q>q_* \:,
\end{array}
\end{cases}
\ee
where \be
\gamma_*=\frac{1-\alpha_*\beta_*}{1-\beta_*}.
\ee
Thus, at the root we have a situation of bifractality: there are two singularities in the distribution function $\mathcal{P}(u)$---which result from the corresponding singularities in the kink $\psi_R(t)$---that control all the moments.

Before closing this subsection, it is worth mentioning that stability conditions corresponding to minimization of the velocity [i.e., analogous to our Eq.~(\ref{beta-frac})] are known to emerge in a broad class of related non-linear problems describing propagation of a front between an unstable and stable phases (here $\psi=1$ and $\psi=0$, respectively). 
The simplest equation of this type, known as Fisher-KPP equation, was introduced by Fisher \cite{fisher1937wave} and by Kolmogorov, Petrovskii, and Piskunov\cite{kolmogorov1937etude} to model the propagation of advantageous genes. Later, it was found that similar problems of traveling waves in reaction-diffusion systems emerge in a variety of contexts, including fluid dynamics, propagation of domain walls in liquid crystals and of combustion fronts, bacterial growth, chemical reactions, etc., see Refs.~\onlinecite{derrida1988polymers,dee1988bistable,van2003front,brunet2015exactly} and references therein. A connection between the problem of Anderson localization at Cayley tree and that of traveling wave propagation was pointed out in Ref.~\onlinecite{monthus2008anderson}.

%\subsection{%Determination of 
%$\xi_{R,r}(t)$ and wave function moments away from the root}
%\label{DetPi}

\subsection{Basic features of $\xi_{R,r}(t)$}
\label{sec:xi}

%%%%%%%%%%%%%%%%%%%%%%
\begin{figure}
\centering
\includegraphics[width=0.5\textwidth]{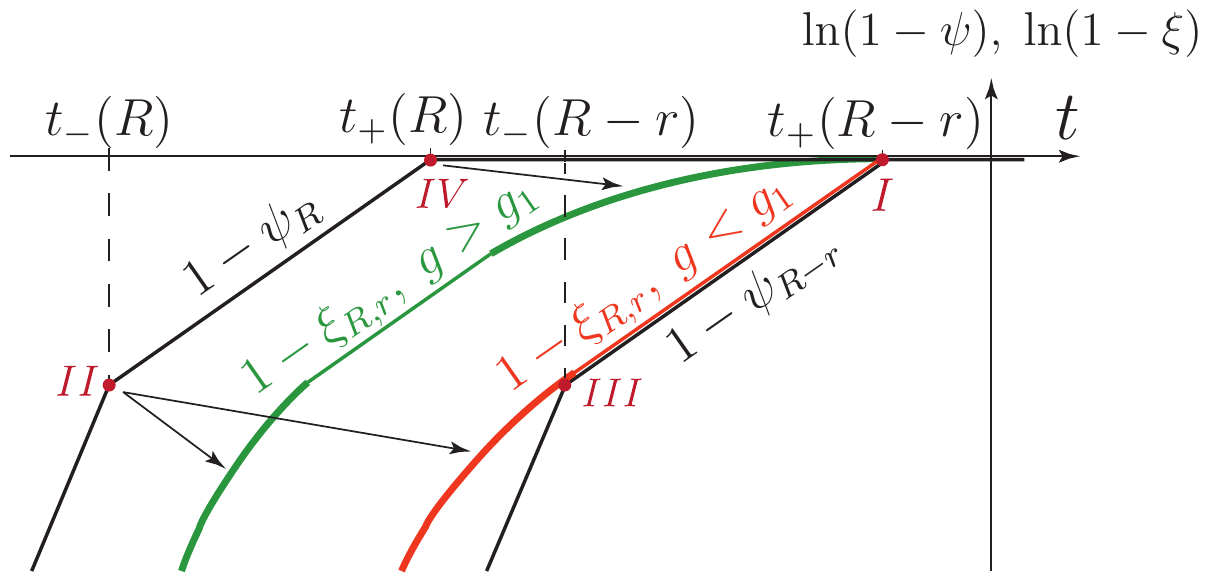}
\caption{Schematic representation of kinks $1- \psi_R(t)$, $1- \psi_{R-r}(t)$ (black lines), and $1- \xi_{R,r}(t)$ (green for $g>g_1$ and red for $g<g_1$) in the limit of large $R$ on the logarithmic scale. The roman numbers I, II, III, and IV mark singularities in  $\psi_R(t)$ and $\psi_{R-r}(t)$ that control the behavior of the wave function moments. The singularities I and III in $\psi_{R-r}(t)$ yield linear-in-$q$ segments of the multifractality spectrum $\tau_q$, Eqs.~(\ref{resI}) and (\ref{resIII}). On the other hand, the singularities II and IV in $\psi_R(t)$, which  develop into curved parts of $\ln[1- \xi_{R,r}(t)]$ (as indicated by arrows), give rise to the non-linear segments (\ref{resII}) and (\ref{resIV}) of the spectrum $\tau_q$.}
\label{fig:schematic_kinks}
\end{figure}
%%%%%%%%%%%%%%%%%%%%%%%

After the reminder of general properties of $\psi_{r}(t)$, which are summarized in Eqs.~(\ref{small-t}), (\ref{large-t}), and (\ref{intermediate-t}) and illustrated in Fig. \ref{fig:schematic_kinks}, and of the resulting bifractality at the root, Eq.~(\ref{qres}), we turn to the analysis of $\xi_{R,r}(t)$ that is needed for evaluating the statistics away from the root.  The function $\xi_{R,r}(t)$ is determined by the recursion (\ref{PiRecursionSimple}) which should be iterated at fixed $R$ from $r=0$ (root) till $r=R$ (leaves, i.e., the boundary of the tree). While this recursive relation is linear with respect to $\xi_{R,r}(t)$, its kernel is itself varying with $r$. Specifically, the behavior of the kernel is governed by that of $\psi_{R-r+1}(t)$ which is the kink studied in Sec.~\ref{DetPsi} but now moving backward (i.e., to the right) with increasing $r$. 

In order to get insight into properties of $\xi_{R,r}(t)$, it is instructive to inspect first the numerical solution of the recurrent relation (\ref{PiRecursionSimple}) which is shown by red lines in Fig. \ref{fig-kinks} for three different values of the coupling constant (belonging to three different regimes). The figure allows us to compare the evolution of $\xi_{R,r}(t)$ with that of $\psi_{R-r}(t)$ (shown by blue lines).

The left panels of Fig.~\ref{fig-kinks}, correspond to the localized phase ($g<g_c$) where the velocity  of the kink $\psi_{R-r}$ [or, more precisely, of its central part, Eq.~(\ref{intermediate-t})] is zero. In this case the kink $\xi_{R,r}(t)$ does not move either, essentially sticking to $\psi_{R-r}(t)$. The middle panels represent a system in the delocalized phase but still at sufficiently strong disorder. In this case, $\psi_{R-r}$ does move with a finite velocity $v_{\beta_*}$, Eq.~(\ref{velocity}), to the right, and $\xi_{R,r}(t)$ essentially follow it. From this point of view, this regime is analogous to the localized phase; the difference between them is in the velocity  $v_{\beta_*}$ being zero in one case and non-zero in the other case. In fact, as will be discussed below, the leftmost tail of $1-\xi_{R,r}(t)$ strongly departs  from $1-\psi_{R-r}(t)$ also in these regimes. This cannot be resolved in Fig. \ref{fig-kinks}, however, since both $\xi_{R,r}(t)$ and $\psi_{R-r}(t)$ are very close to unity there. Finally, the right panels correspond to the part of the delocalized phase with sufficiently weak disorder. Here the function  $\xi_{R,r}(t)$ moves more slowly than $\psi_{R-r}(t)$ and gets strongly distorted. This regime includes the whole ergodic-at-the-root ($g>g_e$) and a part of the non-ergodic-at-the-root ($g<g_e$) domains of the delocalized phase; the only difference between them is that in the ergodic phase the kink in $\psi_{R-r}(t)$ moves with the largest possible velocity $v_{\beta_*} = \ln m$. 

Let us now turn to explanation of these basic properties. Let us recall that the  initial condition for the recursion, Eq. (\ref{xi-initial}), implies that $\xi_{R,r}(t)$ starts at $r=0$ equal to $\psi_R(t)$. Let us first focus on a region of $t$ where $\xi_{R,r}(t)$ moves essentially more slowly that $\psi_{R-r}(t)$, which results in $1 - \xi_{R,r}(t) \gg 1- \psi_{R-r}(t)$. (As we explain below, this is always the case for $t$ that are negative and sufficiently large by absolute value.) In this region, we can approximate $\psi_{R-r+1}(t')$ in the kernel of the integral in Eq.~(\ref{PiRecursionSimple}) by unity. The factor $e^{-e^{t'}}$ can also be replaced by unity in this region. We are then left with a linear recursion relation whose kernel is $L(t-t')$. It is worth mentioning that such kind of recursions conventionally arise in the analysis of one-dimensional chains (that can be formally obtained from the Cayley tree by setting $m=1$), see Refs.~\onlinecite{fyodorov94,mirlin2000statistics}. For brevity, we will term the evolution of $\xi_{R,r}(t)$ with $r$ in the considered region the ``1D evolution''. We now ask the question whether the middle part of the kink $\xi_{R,r}(t)$ with the behavior of the type (\ref{intermediate-t}), i.e., $1-\xi_{R,r}(t) \propto e^{\beta_*t}$, belongs to such a ``1D'' regime. Let us assume that it is the case. Then, this part will be translated to the right with the velocity 
\be 
\label{v-beta-tilde}
\tilde{v}_{\beta_*}=-\frac{\ln\epsilon_{\beta_*}}{\beta_*},
\ee
which can be straightforwardly obtained from Eq.~(\ref{velocity}) by setting $m=1$ and changing the sign. The sign charge is related to the fact that Eq.~(\ref{velocity}) is the velocity of motion of the solution of Eq.~(\ref{PsiRecursionSimple}) to the left, while Eq.~(\ref{v-beta-tilde}) is the velocity of the motion of the solution of the corresponding equation with $m=1$ to the right. Since $\epsilon_{\beta_*} \le 1$, the velocity (\ref{v-beta-tilde}) is non-negative, $ \tilde{v}_{\beta_*} \ge 0$. 

Two scenarios are now possible depending on the relation between the velocities $v_{\beta_*}$ and $\tilde{v}_{\beta_*}$. Note that the velocity $v_{\beta_*}$ is equal to zero at $g=g_c$ and monotonously increases in the interval $g_c < g < g_e$. On the other hand,  the velocity $\tilde{v}_{\beta_*}$ monotonously decreases in the same interval of the coupling $g$, reaching zero at its right border, $g=g_e$  (where $\beta_*=1$ and thus $\epsilon_{\beta_*} = 1$). It follows that there is a point $g_1$ in this interval where the two velocities are equal, which is equivalent to the condition
\be 
m \epsilon_{\beta_*} ^2 = 1.
\ee
Thus,  $\tilde{v}_{\beta_*}<v_{\beta_*}$ for $g>g_1$ and $\tilde{v}_{\beta_*}> v_{\beta_*}$ for $g < g_1$. Let us consider these two situations separately.

In the case of $g<g_1$, the central segment of $\xi_{R,r}(t)$ characterized by the exponent $\beta_*$ would drift according to the 1D regime faster then the corresponding segment in $\psi_{R-r}(t)$. This would, however, violate the underlying assumption of the emergence of the 1D regime. What happens instead is that $\xi_{R,r}(t)$ moves together with  $\psi_{R-r}(t)$. Indeed, linearizing Eq.~(\ref{PiRecursionSimple}) around unity and  making an ansatz
\be
1- \xi_{R,r} = C (1- \psi_{R-r}), \qquad C>0,
\ee
we get an equation for the coefficient $C$,
\be 
1 + \frac{m-1}{C} = \frac{1}{m \epsilon_{\beta_*} ^2},
\ee
which has a positive solution for $g<g_1$. In the localized phase ($g<g_c$) we have $C=1$, i.e., $\xi$ and $\psi$ become asymptotically identical, while in the part $g_c < g < g_1$ of the delocalized phase $C$ satisfies $1 < C < \infty$. The latter implies that  $\xi_{R,r}$ trails $\psi_{R-r}$ by moving with the same velocity and staying a few ($\ln C / \ln m\epsilon_{\beta_*}$) iteration steps behind.  This is exactly what is observed in numerical solution of the recurrence relation for sufficiently small $g$, see middle panels of Fig. \ref{fig-kinks}. We emphasize once more that the conclusion that two kinks move together for $g < g_1$ concerns the segment characterized by the exponent $\beta_*$ but not the region of the largest (by absolute value) negative arguments, $t < t_-$. Indeed, the corresponding part of the function $\psi_r(t)$ is characterized by the exponent $\beta=1$, see Eq.~(\ref{small-t}). Since $\tilde{v}_1 = 0$ and $v_1 = \ln m$, we have $\tilde{v}_1 < v_1 $. Therefore, for the far left asymptotics the 1D regime is applicable: this part of $\xi_{R,r}(t)$ exhibits a slower motion than $\psi_{R-r}(t)$ and gets distorted. The evolution of $1-\xi_{R,r}(t)$ on the logarithmic scale is schematically shown by red line in Fig.\ref{fig:schematic_kinks}.

When $g$ approaches $g_1$ from below, the constants $C$---and thus the distance between the $\xi$ and $\psi$ kinks---diverges, indicating that a new type of behavior should emerge for $g>g_1$. Indeed,
in the case $g>g_1$ the situation is qualitatively different. Now $\tilde{v}_{\beta_*}<v_{\beta_*}$, so that the $\beta_*$ segment of $\xi_{R,r}(t)$ moves with the velocity smaller than that of the corresponding part of $\psi_{R-r}(t)$. This situation is shown in Fig. \ref{fig:schematic_kinks} by the green line. As we will see below, both singularities at $t_-$ and $t_+$ gets smeared under evolution, giving rise to curved parts of $\ln[1- \xi_{R,r}(t)]$  (bold segments of the green line in Fig. \ref{fig:schematic_kinks}). As a result, the whole kink gets strongly distorted. This behavior is manifest in the numerical solution of the recursive relation in the right panels of Fig.~\ref{fig-kinks}. 

Having understood the most salient features of the solution $\xi_{R,r}(t)$ and, in particular, the reason for the emergence of two different types of its behavior in the regimes $g<g_1$ and $g>g_1$, we turn to the quantitative evaluation of the  moments of wave functions. As we show below, the scaling of moments can be analyzed directly, without first evaluating the segments of the solution  $\xi_{R,r}(t)$ represented by curved bold lines in Fig.~\ref{fig:schematic_kinks}.

\subsection{Wave function moments away from the root}
\label{sec:moments-away-root}

\begin{figure}
\centering
\includegraphics[width=0.5\textwidth]{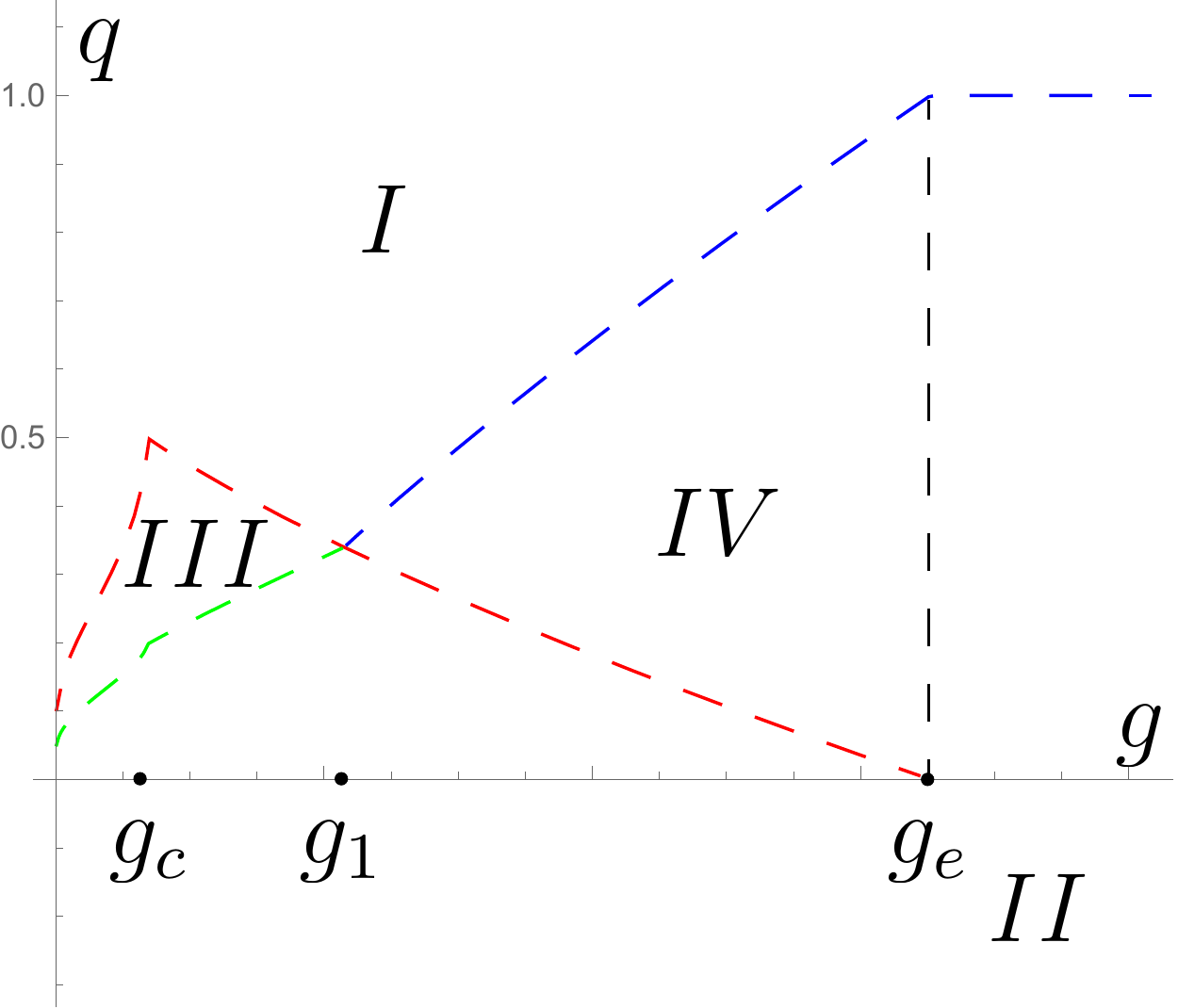}
\caption{``Phase diagram'': regions with different analytical behavior of the multifractal spectrum $\tau_q$ in the parameter plane spanned by $q$ and coupling constant $g$. Characteristic values of the coupling: $g_c$ -- localization transition, $g_e$ separates regions of the delocalized phase with ergodicity and non-ergodicity at the root, $g_1$ separates regions with different types of behavior of  $\xi_{R,r}(t)$, see Sec.~\ref{sec:xi}  and Fig.~\ref{fig:schematic_kinks}. Roman numbers from I to IV label four regimes of the multifractal behavior as discussed in Sec.~\ref{sec:moments-away-root}. The behavior of  $\tau_q$ in each of the regimes originates from a singularity in either $\psi_R$ or $\psi_{R-r}$; these singularities are labeled accordingly in Fig.~\ref{fig:schematic_kinks}. Dashed lines represent borders between the regimes.  Red line: $q_*(g)$, green line: $Q_*(g)$, blue line: $Q(g)$. 
The phase boundaries in the plot have been calculated for the branching number $m=2$, in which case $g_c \simeq 0.068$, $g_1 \simeq 0.215$, and $g_e \simeq 0.65$. The phase diagram has qualitatively the same form for other values of $m$ as well. Multifractality spectrum $\tau_q$ in each of the intervals of the values of $g$  is shown schematically in Fig. \ref{fig_taus}.
}
\label{pd}
\end{figure}

\begin{figure}
\centering
\includegraphics[width=0.5\textwidth]{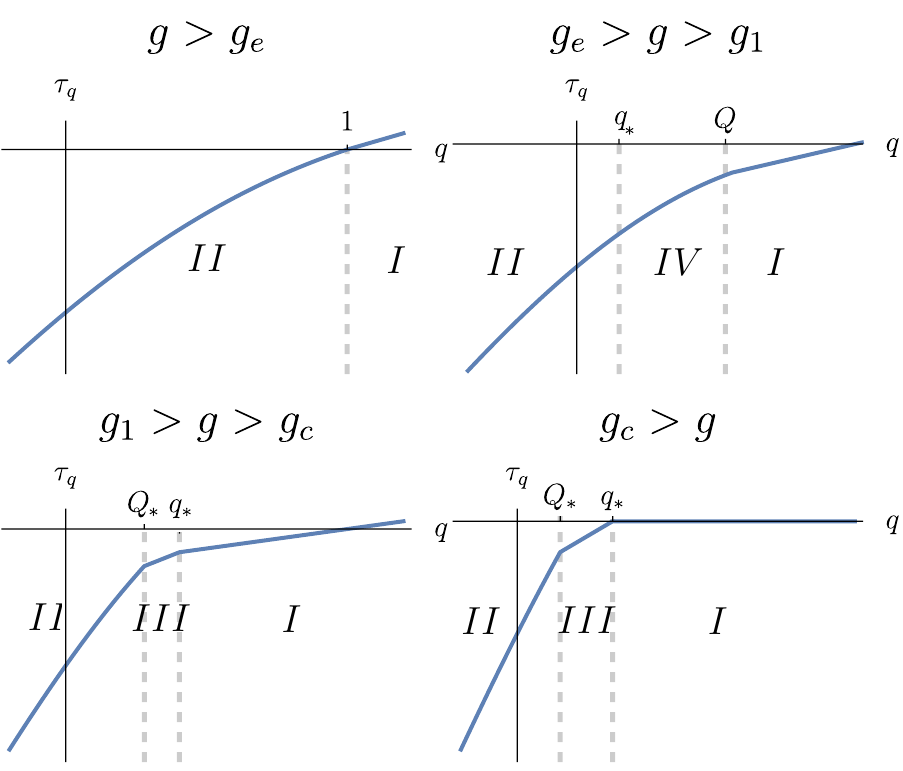}
\caption{Schematic behavior of the multifractality spectrum $\tau_q$ (for a generic value of $s$, i.e., $0<s<1$) in each of the intervals of the values of the $\sigma$ model coupling $g$ from the phase diagram of Fig. \ref{pd}.  The labels I, II, III, and IV refer to four different regimes of the multifractal behavior, see Fig.\ref{pd} and Sec.~\ref{sec:phase-diagram}. See also right panels of Figs.~\ref{qdep5}, \ref{qdep8}, \ref{qdep14}, and \ref{qdep18} below for numerical plots of the evolution of $\tau_q$ with $s$ for representative values of $g$ in each of these intervals.}
\label{fig_taus}
\end{figure}

At the root, the scaling of a moment $P_q$ is controlled by one of the singular points of $\psi_R(t)$, i.e., $t_+(R)$ or $t_-(R)$, depending on the value of $q$. This yield two different analytical formulas for the fractal exponents $\tau_q$ in the regions $q>q_*$ and $q<q_*$, see Eq. (\ref{qres}). As we show below, away from the root four different types of analytical behavior of $\tau_q$ emerge that we label by roman numbers from I to IV. More specifically, for any given value of the coupling $g$ (i.e., of disorder), the axis of $q$ is divided in three regions (that we will term regions of high, intermediate, and low moments) with distinct analytical forms of $\tau_q$. The analytical forms of $\tau_q$ in the regions of high and low moments are the same for all $g$; we label them by I and II, respectively. The behavior of intermediate moments is different for $g<g_1$ and $g>g_1$ and is labeled by III and IV, respectively. The emerging ``phase diagram'' of the multifractal behavior in the parameter plane spanned by $g$ and $q$ is shown in Fig.~\ref{pd}.
 The evolution of the multifractality spectrum $\tau_q$ with $g$ is illustrated schematically in Fig. \ref{fig_taus}.
%%%%%%%%%%%%%%%%%%%
%%%%%%%%%%%%%%%%%%%%

%%%%%%%%%%%%%%%%%%%
%%%%%%%%%%%%%%%%%%%%

%\subsubsection{Large positive and large negative $q$.} 

\subsubsection{High moments:  Region I} 
\label{sec:high}

The distribution function is given by Eq.~(\ref{Pu}) and thus determined, in view of of Eqs.~(\ref{Yar}) and (\ref{Phi}), by the product $\psi_{R-r}^m(t) \xi_{R,r}(t)$. Since in all the regimes $1- \xi_{R,r}(t)$ is either of the same order as or much larger than $1- \psi_{R-r}(t)$, the scaling behavior will be always determined by $\xi_{R,r}(t)$. For large system, $N\gg 1$, the leading behavior of each moment $q$ is controlled, by virtue of saddle-point approximation, by a certain point on the distribution function, and thus by a certain value of the argument $t$ of the function $\xi_{R,r}(t)$. With increasing $q$ this points moves to the right: the larger the moment $q$ is, the larger are values of the wave function amplitude that dominate it. This motion of the characteristic $t$ to the right is limited by $t_+(R-r)$ since $\psi_{R-r}$ and thus also $\xi_{R,r}(t)$ are negligibly small for larger $t$. 
As a result, sufficiently high moments will be determined by the point $t_+(R-r)$, i.e., by the values of the wave function intensity $u \sim u_+(R-r) \equiv e^{t_+(R-r)}$. 

The fact that high moments are controlled by the singularity at $t_+(R-r)$ is particularly clear for $g < g_1$. Indeed, in this case, $\xi_{R,r}(t)$ can be simply replaced by  $\psi_{R-r}(t)$ for the analysis of scaling of sufficiently high moments, see red line in Fig.~\ref{fig:schematic_kinks}. This yields the distribution function 
\bea
\label{distr-high-moments}
{\cal P}(u) &\sim& N^{-1+(1-s) \alpha_*\beta_*} u^{\beta_*-2}, \nonumber \\ && \qquad   u_-(R-r) < u < u_+(R-r),
\eea
where we have introduced the parameter $s$, 
\be
s=r/R \,,
\label{s}
\ee 
which is the distance from the observation point to the root divided by the ``radius'' of the tree. The borders of the power-law behavior (\ref{distr-high-moments}) of the distribution function are 
\bea
u_-(R-r) &=& e^{t_-(R-r)} \sim N^{-\gamma_*(1-s)} \,; 
\label{u-minus}  \\
u_+(R-r) &=& e^{t_+(R-r)} \sim N^{-\alpha_*(1-s)} \,.
\label{u-minus}
\eea
It follows that the moments $P_q$ with $q > q_*$, with $q_*$ given by Eq.~(\ref{qstar}), are determined by the right border $u_+(R-r)$ of the interval and scale  according to Eq.~(\ref{Pq-scaling}) with the exponents $\tau_q$ given by
\be
\label{resI}
\tau_q=(q-1)(1-s)\alpha_* \,.
\ee 

Let us now turn to the range of weaker disorder, $g > g_1$. In this case, the $\beta_*$ segment of $\xi_{R,r}(t)$ moves with a velocity smaller than that of $\psi_{R-r}$, see Sec.~\ref{sec:xi}. Nevertheless, the front of the kink $\xi_{R,r}$ still reaches the frontal point $t_+(R-r)$ of $\psi_{R-r}$, see green line in Fig.~\ref{fig:schematic_kinks}. This will be confirmed below by the calculation of moments with intermediate $q$ that are controlled by the right bold segment of the green line in Fig.~\ref{fig:schematic_kinks} (region IV). Thus, Eq.~(\ref{resI}) for the scaling exponents of high moments applies at $g>g_1$ as well.   The determination of lower border  of the range of $q$ for which the scaling (\ref{resI}) applies at $g>g_1$ requires, however, an additional analysis. We will find this border below after evaluating $\tau_q$ in the range of intermediate $q$ (region IV).

\subsubsection{Low moments:  Region II}
\label{sec:low}

Next, we analyze the scaling of  low moments $P_q$ (region II), i.e., those with $q$ smaller than certain value (to be determined). We will see that the upper border of region II is positive, so that all negative $q$ belong to this region.  In full analogy with the high moments determined by the right tail of the function $\xi_{R,r}(t)$, see Sec.~\ref{sec:high}, the low moments are determined by its left tail. 

Since the distribution function, Eq. (\ref{Pu}), is expressed as a second derivative, it is useful to intergate twice by part in the expression for the moments, 
\be
P_{q,r} \sim   \int u^q \partial^2_u \xi_{R,r}(t)du \sim \int e^{(q-1)t}\xi_{R,r}(t)dt.
\ee
For the convenience of the analysis, we indicated here by a second subscript that the moment refers to wave functions at the distance $r$.

Now we use the recurrent relation (\ref{PiRecursionSimple}) for $\xi_{R,r}(t)$. Since low moments are dominated by the region where $1-\xi_{R,r}(t) \gg 1- \psi_{R-r}(t)$ (the situation we have termed the ``1D regime''), we can replace $\psi_{R-r+1}(t)$ by unity in the recurrence relation, see the discussion in Sec.~\ref{sec:xi}.  This yields
\be
P_{q,r} \sim \int \int e^{(q-1)t}L(t-t')\xi_{R,r-1}(t')dtdt'.
\ee
Since $L(t) = e^t L(-t)$, a function $e^{\beta t}$ is a left eigenfunction of the integral operator $\hat L$ with the eigenvalue $\epsilon_{1-\beta}$. 
Thus, we come to the following recursion for the moments:
\be
\label{PRec}
P_{q,r}=\epsilon_q P_{q,r-1}.
\ee
Iterating this recursion, we get a relation between the low moments at a point $r$ on the tree and at the root ($r=0$):
\be
\label{PRec-iterated}
P_{q,r}=\epsilon_q^r P_{q,0}.
\ee
On the other hand, the scaling of low moments $P_{q,0}$ at the root is given by the first line of Eq.~(\ref{qres}). Substituting this in Eq.~(\ref{PRec-iterated}), we find the fractal exponents $\tau_q$ at the point $r$,
\be
\label{resII}
\tau_q=q\gamma_*-1-s\frac{\ln\epsilon_{q}}{\ln m} \,,
\ee
where again introduced the variable $s$, Eq.~(\ref{s}), parametrizing the position of the observation point on the lattice. 

\subsubsection{Intermediate moments at $g<g_1$: Region III} 
\label{sec:intermediate-small-g}

We turn now to the behaviour of moments with intermediate $q$. This analysis should be done separately for the cases of stronger ($g<g_1$) and weaker ($g>g_1$) disorder.  We begin with the regime of relatively strong disorder,  $g<g_1$ (region III). As discussed above, in this case the segment of $\xi_{R,r}(t)$ characterized by the exponent $\beta_*$ follows the function $\psi_{R-r}$. This segment is limited by two singularity points, $t_+(R-r)$ and $t_-(R-r)$. The first of them determines the scaling of high moments (region I), see Sec.~\ref{sec:high} above. The second singularity, $t_-(R-r)$, which yields the lower border of the power-law regime (\ref{distr-high-moments}), determines  the scaling of intermediate moments. Using Eq.~(\ref{distr-high-moments}) for the distribution function, we find the following scaling exponents for the moments that are determined by  the lower border (\ref{u-minus}) of the power-law regime: 
\be
\label{resIII}
\tau_q=\left(q\gamma_*-1\right)(1-s).
\ee

The region III is located between the region I of high moments and the region II of low moments that have been discussed above. 
Since the singularities I and III giving rise to the corresponding regions are separated by a domain of the power-law behavior of the distribution function, Eq.~(\ref{distr-high-moments}), the border between these regimes is $q=q_*$ given by Eq. (\ref{qstar}). Indeed, it is easy to check that  Eqs.~(\ref{resI}) and (\ref{resIII}) match at that point. To find the border between the regimes III and II, we equate the corresponding formulas for $\tau_q$. This yields 
$q=Q_*$ defined implicitly via the equation
\be
\label{Bstar}
Q_*\gamma_*-1 = \frac{\ln \epsilon_{Q_*}}{\ln m}.
\ee

\subsubsection{Intermediate moments at $g>g_1$: Region IV}

Finally, we turn to the analysis of the intermediate region IV that separates the regions of high and low moments at weaker disorder ($g>g_1$). 
We recall that for $g>g_1$ almost the whole function $\xi_{R,r}(t)$ stays well behind $\psi_{R-r}(t)$ (apart from its far right tail that determines the scaling of high moments, region I), see the green line in Fig.~\ref{fig:schematic_kinks}. Therefore, not only low moments but also intermediate moments will be determined in this case 
by the recurrence relation (\ref{PiRecursionSimple}) for $\xi_{R,r}(t)$ in which  $\psi_{R-r+1}(t)$ is replaced by unity. In other words, the analysis of Sec.~\ref{sec:low} leading to Eq.~(\ref{PRec-iterated}) fully applies in this case as well. In Sec.~\ref{sec:low} we combined Eq.~(\ref{PRec-iterated}) with the first line of Eq.~(\ref{qres}) representing low moments at the root. Now we have to apply this also to higher moments that are dominated at the root by the singularity $t_+(R)$, see Fig.~\ref{fig:schematic_kinks}. The scaling of these moments is given by the second line of Eq.~(\ref{qres}). Substituting it in  Eq.~(\ref{PRec-iterated}), we find
\be
\label{resIV}
\tau_q=(q-1)\alpha_* - s\frac{\ln\epsilon_{q}}{\ln m}.
\ee

We determine now the borders between the intermediate region IV and the regions I (high momenta) and II (low momenta). It is straightforward to check that regions IV and II match at $q=q_*$ and 
matching of IV and I happens at $q=Q$ defined by the equation
\be
\label{Q}
\alpha_*(Q-1)=\frac{\ln\epsilon_Q}{\ln m}.
\ee

\subsubsection{Phase diagram}
\label{sec:phase-diagram}

We have thus found four different regimes I--IV of behavior  of the multifractality spectrum $\tau_q$. 
As is clear from the derivation, each of the four obtained regimes originates from a singularity in either $\psi_R$ or $\psi_{R-r}$; these singularities are labeled accordingly in Fig.~\ref{fig:schematic_kinks}. Let us summarize the obtained results for $\tau_q$ and their origin: 
\begin{itemize}
\item Region I:  High moments, determined by an upper cutoff of the distribution function, i.e., by $t_+(R-r)$ in Fig.~\ref{fig:schematic_kinks}:
$\tau_q=(q-1)(1-s)\alpha_*$.
\item Region II: Low moments, determined by the lowest values of the wave function. The corresponding behavior of $\tau_q$ originates in the singularity $t_-(R)$ of $\psi_R$ evolved according to the 1D recurrence, with the result 
$\tau_q=q\gamma_*-1-s\ln\epsilon_{q} / \ln m$.
\item Region III:  Intermediate moments at high disorder, $g<g_1$,  determined by  the lower limit of the power-law behavior in the distribution function, i.e. by $t_-(R-r)$, yielding
$\tau_q=\left(q\gamma_*-1\right)(1-s)$.
\item Region IV:  Intermediate moments at low disorder, $g>g_1$. Behavior of $\tau_q$ in this region originates in the singularity $t_+(R)$ of $\psi_R$ evolved according to the 1D recurrence, which gives
$\tau_q=(q-1)\alpha_*-s\ln\epsilon_{q}/ \ln m$.
\end{itemize}
Here $\alpha_*$, $\gamma_*$, and $\epsilon_q$ are functions of the coupling $g$ as defined in Sec.~\ref{SectionAnalytical}. 

The corresponding regions in the $(g,q)$ parameter plane are shown in  Fig.~\ref{pd}. The boundary between the domains are represented in the figure by dashed lines. Specifically, the red line is $q_*(g)=1-\beta_*(g)$ at $0<g<g_e$, while the green and the blue lines are $Q_*$ at $0<g<g_1$ and $Q$ at $g_1<g$, respectively, which are defined implicitly via Eqs.~(\ref{Bstar}) and (\ref{Q}). The phase boundaries in Fig.~\ref{pd} have been calculated for a Cayley tree with the branching number $m=2$; in this case the characteristic values of the coupling $g$ are $g_c \simeq 0.068$, $g_1 \simeq 0.215$, and $g_e \simeq 0.65$. The phase diagram retains qualitatively the same form for other values of $m$.

In the next Section we will compare these analytical findings (obtained for the $\sigma$ model corresponding to the $n$-orbital model in the large-$n$ limit) with the results of direct numerical simulations of the $n=1$ Anderson model. 

\section{Wave function multifractality: Numerical results}
\label{numerics}

We have studied numerically the wave function moments $P_q$ given by Eq.~(\ref{Pq})  by performing the exact diagonalization of the $n=1$ Anderson model on a Cayley tree with the branching number $m=2$.  To determine the mulitfractal exponents $\tau_q$, we have analyzed the scaling of the moments with the number of sites $N$, see Eq.~(\ref{Pq-scaling}). We have explored systems of radius $R$ in the range from $R = 7$ to $R = 14$, considering states in the middle of the spectrum ($1/8$ eigenvalues around the center of the band, $\varepsilon=0$). 
The averaging was performed over $10^6$ to $10^5$ wavefunctions for the system size $R$ from 7 to 14, respectively. For each wave function, we have additionally averaged over all points of the lattice located at a given distance $r$ from the root. 

In the figures presented in this Section, the obtained numerical results for the $n=1$ model are compared with the analytical findings of Sec.~\ref{SectionAnalytical} for the $n\gg 1$ model (i.e., the $\sigma$ model). Since the $\sigma$-model coupling constant $g$ scales with disorder $W$ as $g \propto W^{-2}$  [see the text below Eq.~(\ref{action})], we establish a correspondence according to $W \longrightarrow g = g_c (W_c/W)^2$.  This ensures that the Anderson-transition critical point of both models, $W_c$ and $g_c$ (equal to $17.5$ and $0.068$, respectively, for $m=2$), are mapped exactly onto each other. 

We find that the analytical and numerical results are in very good agreement: although the $\sigma$ model is, strictly speaking, valid only in the $n\to \infty$ limit, it captures correctly essential features of the $n=1$ model.  The numerically obtained moments (\ref{Pq}) of the wave functions indeed show a power-law scaling with $N$, as predicted by Eq.~(\ref{Pq-scaling}), which allows us to extract the multifractal exponents $\tau_q$. Our numerical results confirm that the proper way to define the position-dependent multifractal spectrum is to keep the ratio $s=r/R$ fixed in the thermodynamic limit $N\to\infty$. The multifractal exponents $\tau_q$ depend on the disorder strength $W$ and on the position $s$ on the Cayley tree.  The qualitative evolution of the multifractal spectrum with $W$ and $s$ is in perfect correspondence with the analytical results for the $n\gg 1$ model. Furthermore, the analytical and numerical values of $\tau_q$ turn out to be quite close quantitatively. 

\begin{figure}[htp]
\centering
\includegraphics[width=.5\textwidth]{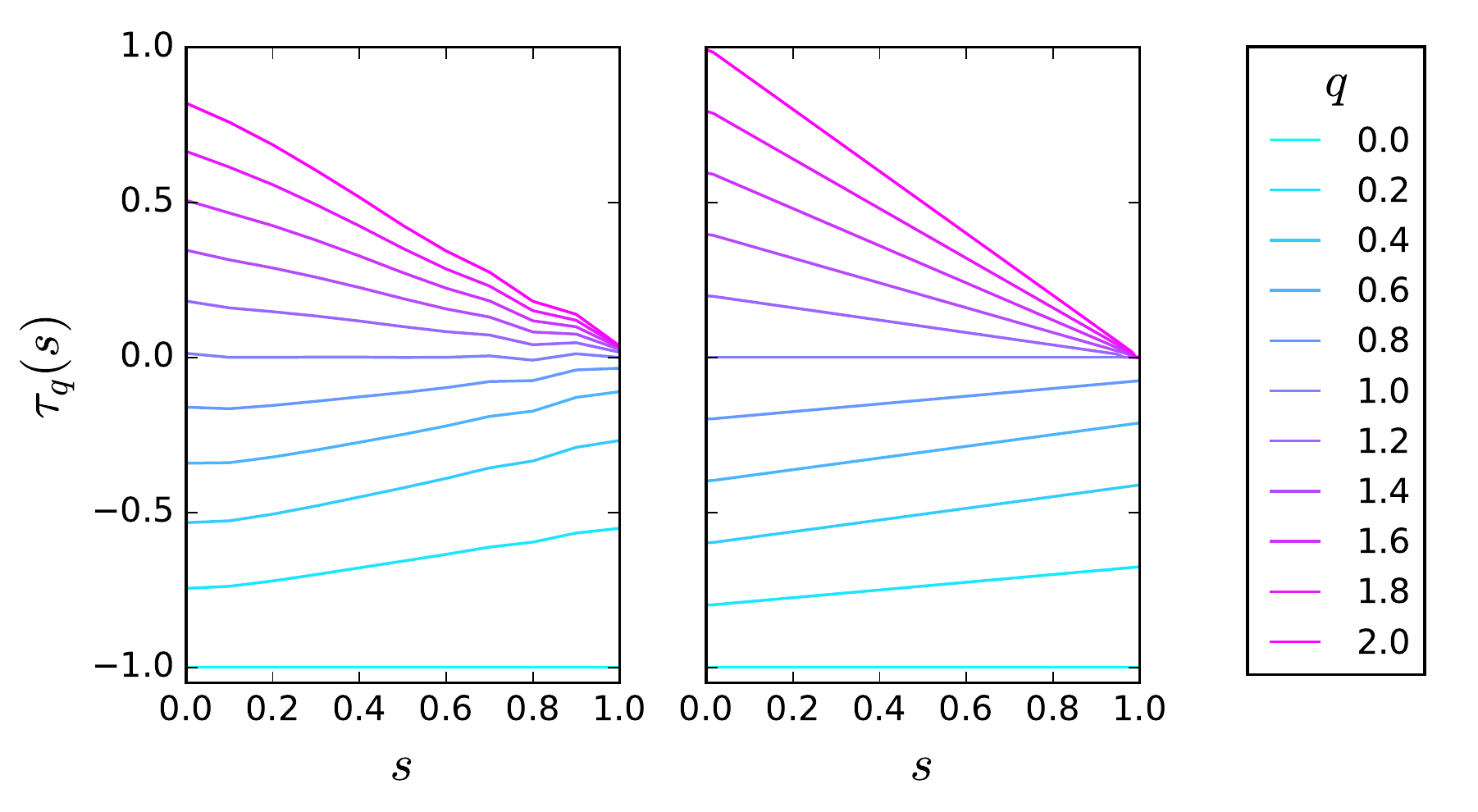}
\caption{Dependence of the fractal exponents $\tau_q$  on the position $s=r/R$ on the Cayley tree with branching number $m=2$ for various $q$ at disorder strength $W=5$. {\it Left:} numerical results for the $n=1$ Anderson model; {\it right:}  analytical results for the $\sigma$ model at the corresponding coupling $g = 0.83$.
}
\label{rdep}
\end{figure}

We begin by showing in Fig.~\ref{rdep} the dependence of the fractal exponents $\tau_q$ on the position $s$ for a relatively weak disorder, $W=5$, and for various $q$ in the range between 0 and 2. For comparison, we show in the right panel the analytical results for the corresponding value of the $\sigma$-model coupling, $g = 0.83$ (which is slightly above $g_e$). The numerical results confirm the linear dependence of the exponents $\tau_q$ on $s$. The observed deviations from the linear dependence (an oscillatory structure near $s=1$ and a rounding near $s=0$) are attributed to finite-size effects.The linear dependence of $\tau_q$ on $s$ is supported also by numerical data for other values of disorder $W$. The finite-size oscillatory corrections near $s=1$ become much weaker with increasing disorder. 

%%%%%%%%%%%%%%%%%%%%%%
%%%%%%%%%%%%%%%%%%%

\begin{figure}[htp]
\centering
\includegraphics[width=.5\textwidth]{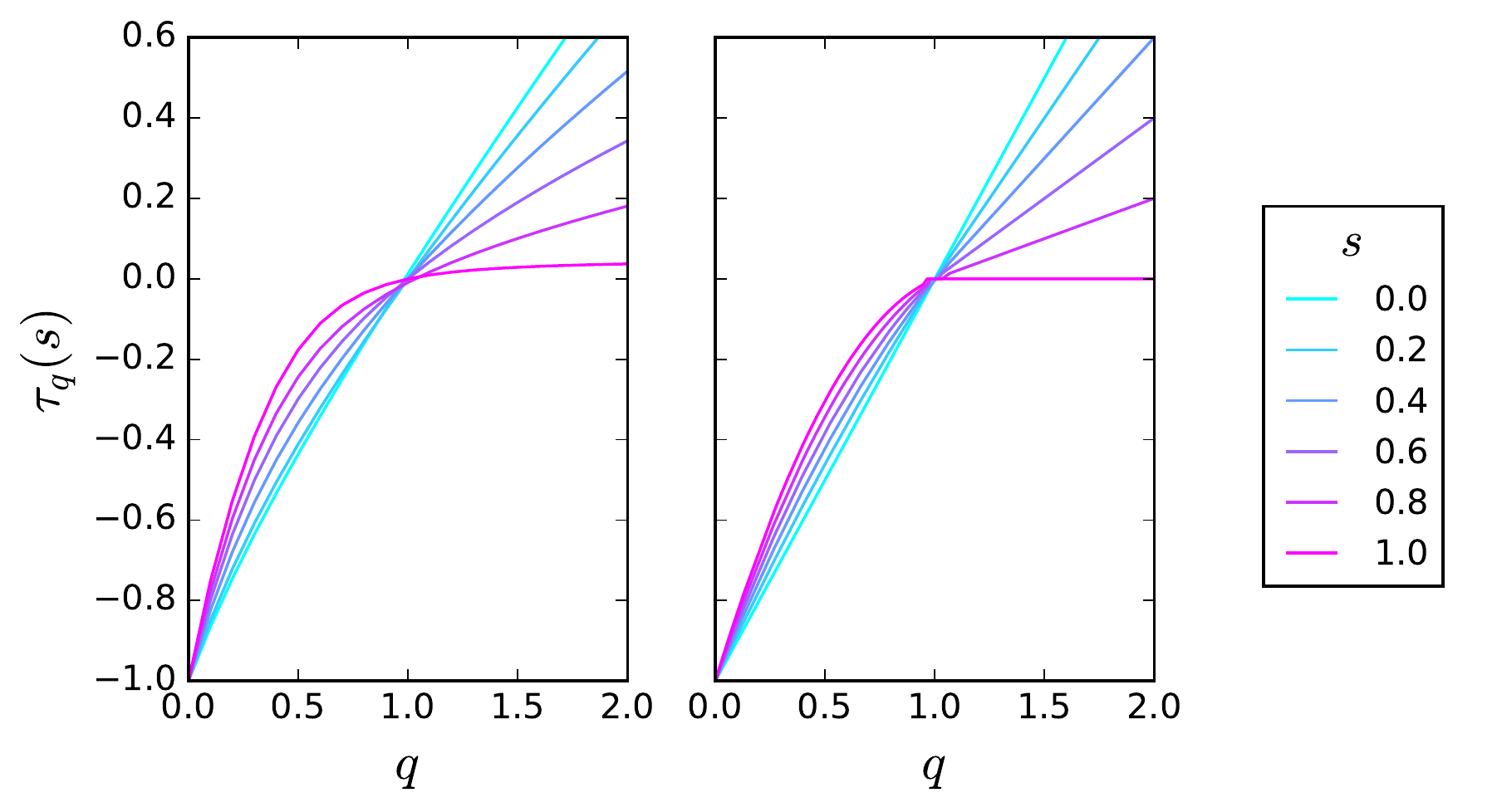}
\caption{Multifractality spectrum $\tau_q$ at various positions $s$ on the tree  for disorder $W=5$. {\it Left:} numerical results for the $n=1$ Anderson model; {\it right:}  analytical results for the $\sigma$ model at the corresponding coupling $g = 0.83$.
 }
\label{qdep5}
\end{figure}

\begin{figure}[htp]
\centering
\includegraphics[width=.5\textwidth]{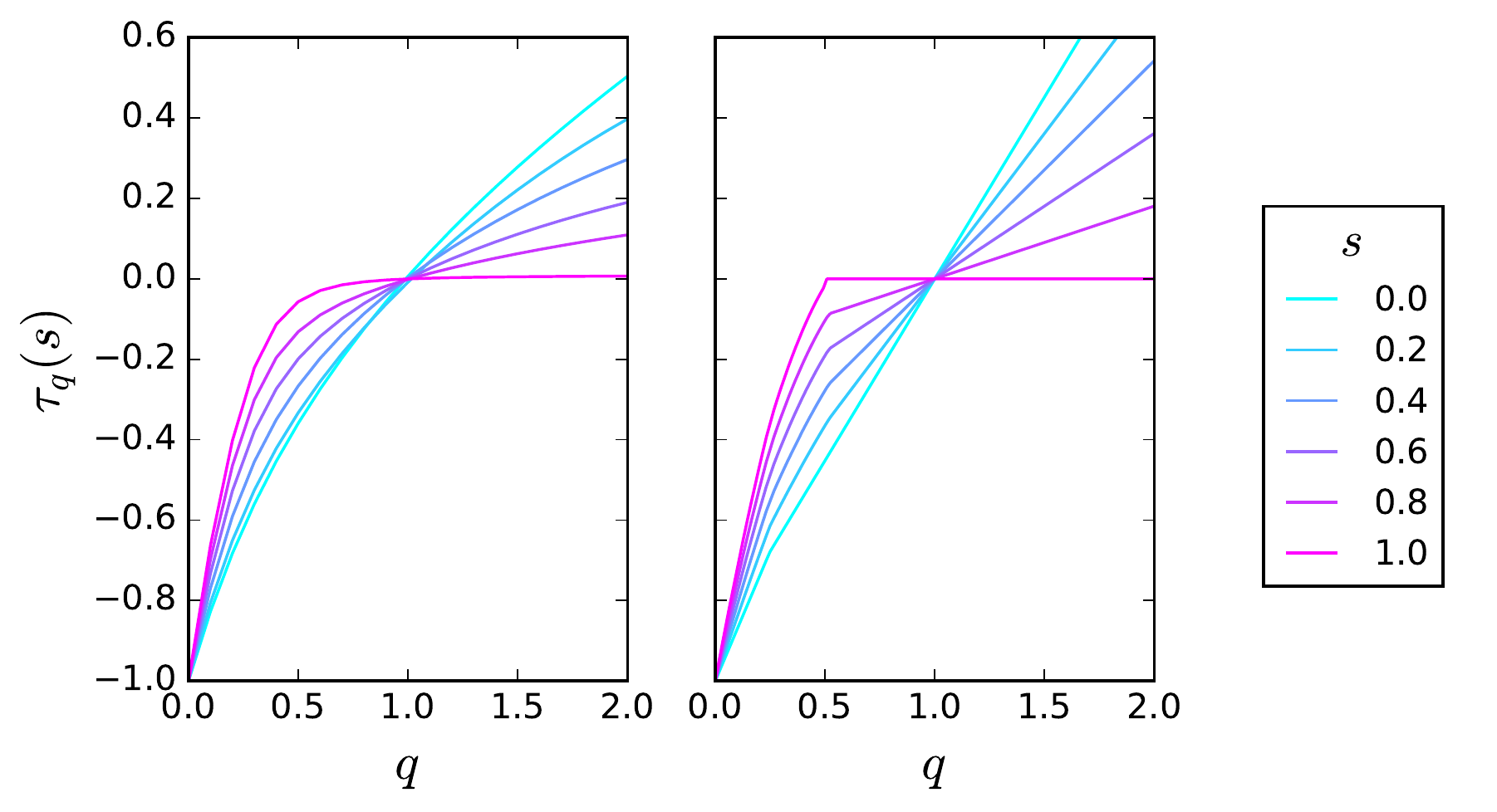}
\caption{Multifractality spectrum $\tau_q$ at various positions $s$ on the tree for disorder $W=8$. {\it Left:} numerical results for the $n=1$ Anderson model; {\it right:}  analytical results for the $\sigma$ model at the corresponding coupling $g = 0.33$.}
\label{qdep8}
\end{figure}

\begin{figure}[htp]
\centering
\includegraphics[width=.5\textwidth]{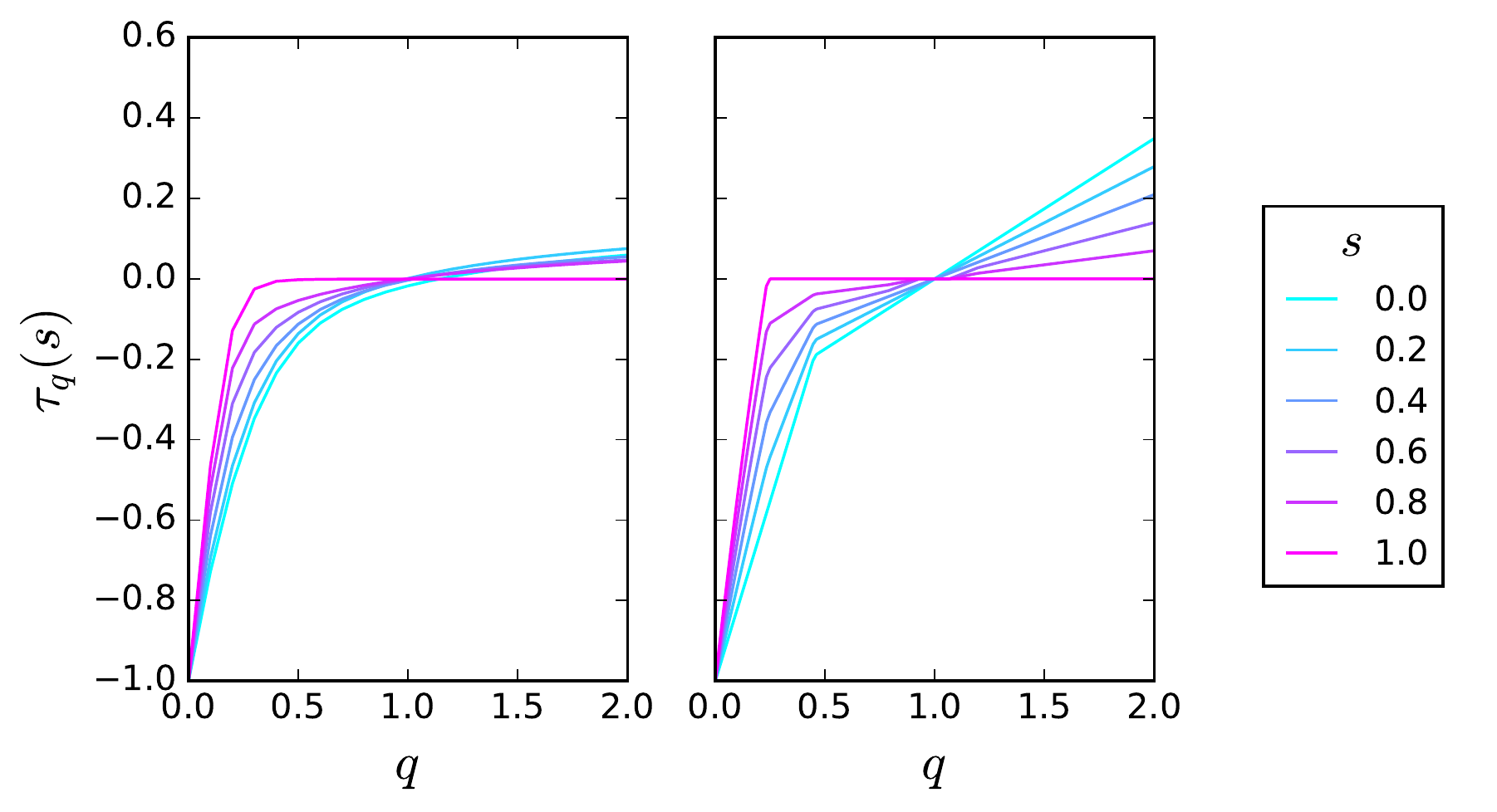}
\caption{Multifractality spectrum $\tau_q$ at various positions $s$ on the tree for disorder $W=14$. {\it Left:} numerical results for the $n=1$ Anderson model; {\it right:}  analytical results for the $\sigma$ model at the corresponding coupling $g = 0.11$.}
\label{qdep14}
\end{figure}

\begin{figure}[htp]
\centering
\includegraphics[width=.5\textwidth]{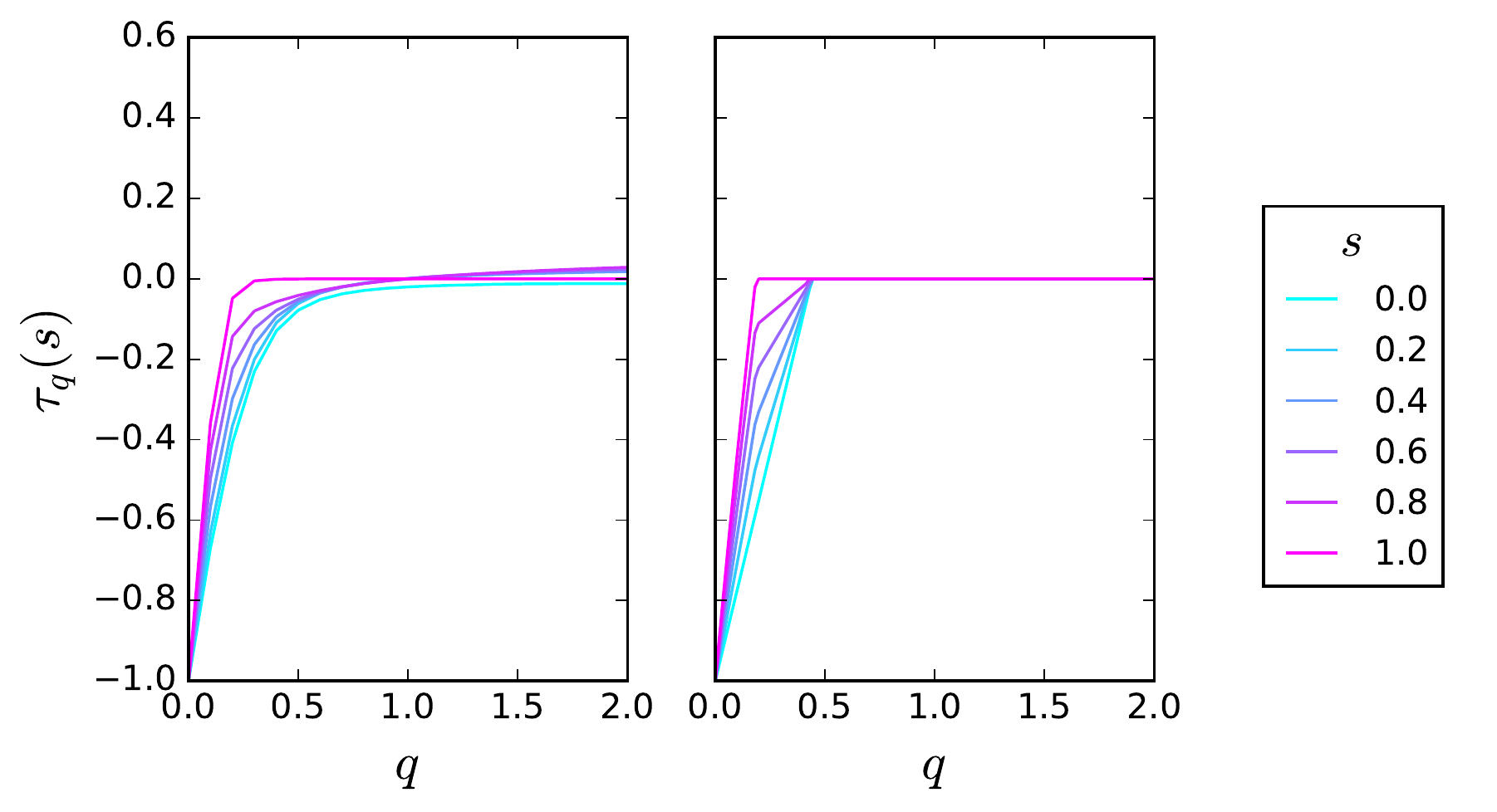}
\caption{Multifractality spectrum $\tau_q$ at various positions $s$ on the tree  for disorder $W=18$. {\it Left:} numerical results for the $n=1$ Anderson model; {\it right:}  analytical results for the $\sigma$ model at the corresponding coupling $g = 0.064$.}
\label{qdep18}
\end{figure}

In Fig.~\ref{qdep5} the data of Fig.~\ref{rdep} (i.e., for $W=5$) are replotted in the way that allows us to show the multiifractal spectrum $\tau_q$ for each value of $s$. Figures \ref{qdep8}, \ref{qdep14}, \ref{qdep18} represent analogous plots for other representative values of disorder. Specifically, in Fig.~\ref{qdep8} we show numerical multifractal spectra for an intermediate disorder $W=8$ and analytical results for the corresponding value of the $\sigma$ model coupling $g = 0.33$ (which belongs to the interval $ g_1 < g < g_e$). Further, in Fig.~\ref{qdep14} results for stronger disorder $W=14$ (close to the Anderson transition but still on the delocalized side) are presented, along with the $\sigma$-model results for the corresponding coupling $g \simeq 0.11$ belonging to the range $g_c < g < g_1$. Finally, Fig.~\ref{qdep18} shows data for the localized side of the transition ($W=18$ and the corresponding $\sigma$-model coupling $g = 0.064$). 

%%%%%%%%%%%%%%%%%%%%%%

%%%%%%%%%%%%%%%%%%%%%%

Thus, the four figures \ref{qdep5}, \ref{qdep8}, \ref{qdep14}, and \ref{qdep18} represent the evolution of the multifractal spectrum $\tau_q$ on a Cayley tree from the root ($s=0$) to the leaves ($s=1$) and from weak to strong disorder. A comparison of left and right panels of these figures shows a remarkable similarity in gross features of this evolution between the $n=1$ numerical results and the $n\gg 1$ theory. 
Let us discuss in more detail several key properties of the multifractal spectra $\tau_q$. One of them, the linear dependence on $s$, has already been emphasized above. Another important feature is the ``localized'' character of the spectrum at the boundary ($s=1$) for any disorder $W$, in the sense that $\tau_q \equiv 0$ for all $q$ exceeding a certain value ($Q$ or $Q_*$ in notations introduced in Sec.~\ref{SectionAnalytical}), see Eqs.~\ref{resI} and \ref{resIII}. Our numerical results indeed show very clearly this feature  (up to finite-size corrections at weak disorder, $W=5$, that have been already mentioned above). This result is in full consistency with the pure-point character of spectrum of random Schr\"odinger operators on canopy graphs that has been mentioned in  Sec.~\ref{SectionIntro}. A further interesting property is that even in the localized phase the spectrum $\epsilon_q$ characterizing eigenfunction moments retains its multifractal character for sufficiently low $q$. This holds in the regions II and IV of the phase diagram for all $s$, as well as in the region III for $s \ne 1$.

%%%%%%%%%%%%%%%%%%%%%%
\begin{figure}[htp]
\centering
\includegraphics[width=.5\textwidth]{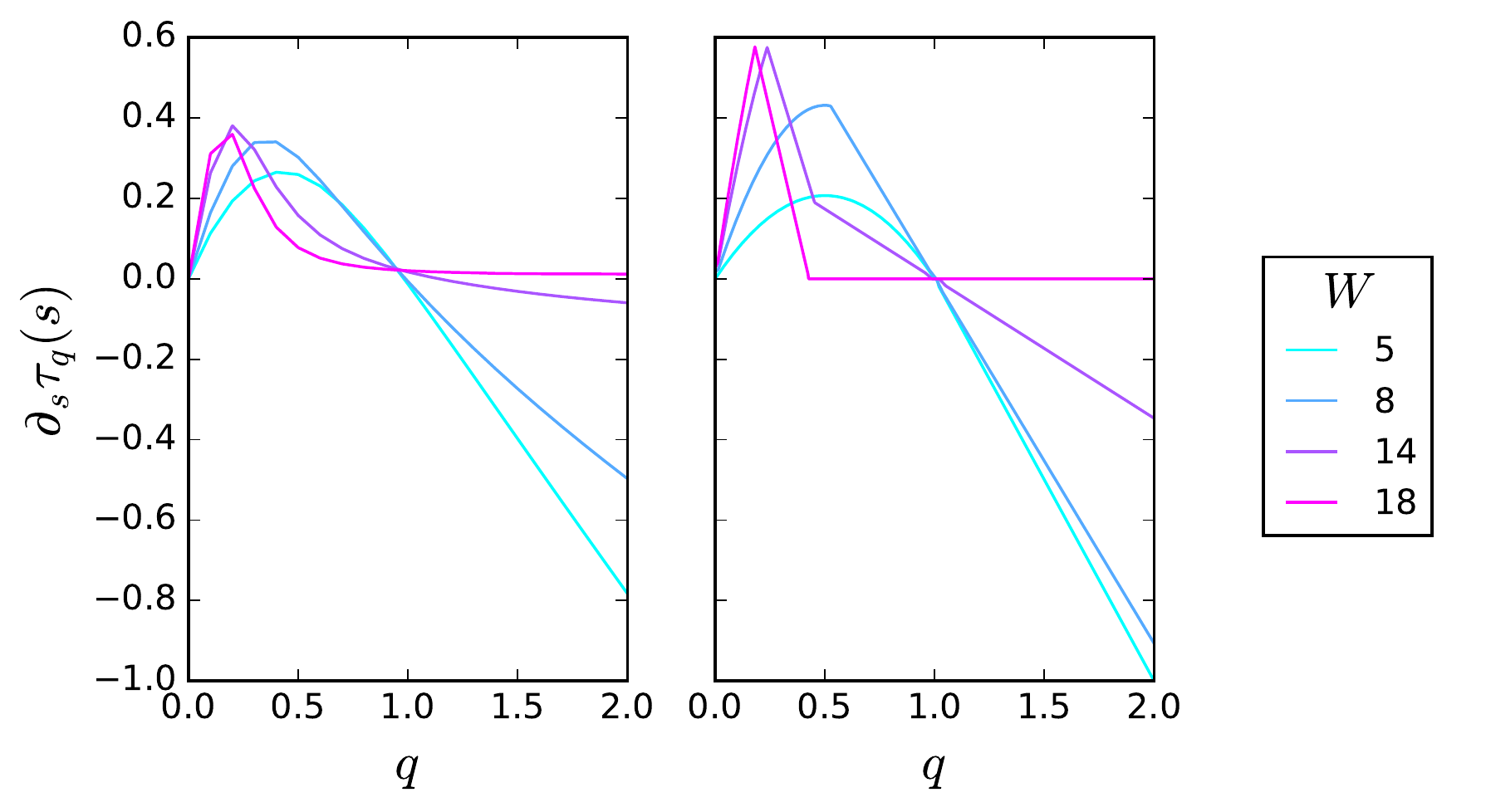}
\caption{ Slope $\partial_s\tau_q(s)$ as a function of $q$, for disorder strength $W=5,\;8,\;14,\;18$.  {\it Left:} numerical results for the $n=1$ Anderson model; {\it right:}  analytical results for the $\sigma$ model at the corresponding couplings ($g=0.83$, 0.33, 0.11, and 0.064, respectively).}
\label{dt}
\end{figure}
%%%%%%%%%%%%%%%%%%%%%%%

Finally, in Fig. \ref{dt} we show the slope $\partial_s \tau_q$ of the linear $s$-dependence of $\tau_q$ (which can be alternatively defined as $\tau_q(s=1) - \tau_q(s=0)$) as a function of $q$ for all four values of disorder, $W=5$, 8, 14, and 18. For the $\sigma$ model, this slope is given by $-\ln \epsilon_q / \ln m$ in the regions II and IV, $(1-q)\alpha_*$ in the region I, and $1-q\gamma_*$ in the region III. Thus, the behavior of this slope at low $q$ (regions II and IV) yields directly the spectrum of eigenvalues $\epsilon_q$ of the integral operator $L(t)$. Interestingly, the shape of the $q$-dependence of the slope is qualitatively different for $g > g_1$ (where it is concave) and $g < g_1$ (where it is neither concave nor convex). The numerical results for the $n=1$ model agree very well with these $\sigma$-model analytical findings.

%%%%%%%%%%%%%%%%%%%%%%%%%%%%%%%%
\begin{figure}
\centering
\includegraphics[width=0.4\textwidth]{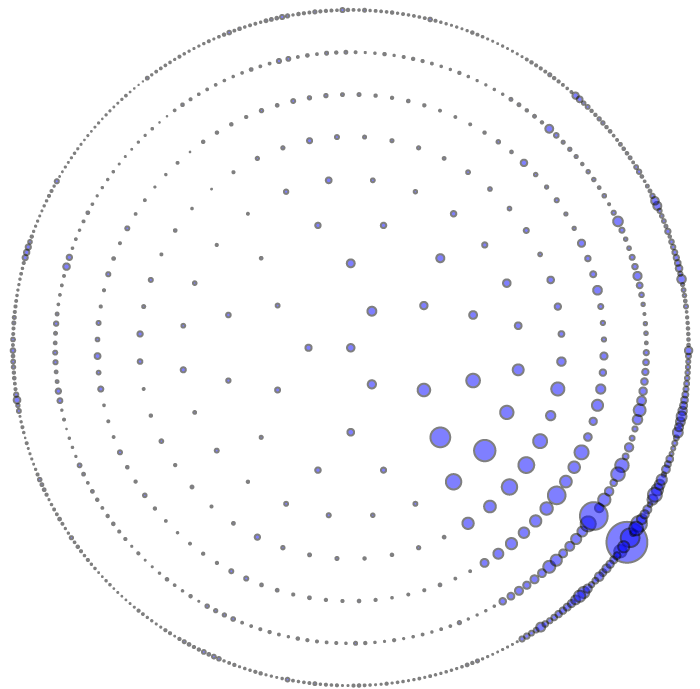}
\label{wfun}
\caption{Typical spatial image of an eigenfunction on a Cayley tree with 8 generations for disorder $W=10$. The area of each circle is proportional to the  absolute value $|\psi|$ of the wave function amplitude on the corresponding site.}
\label{wfun}
\end{figure}
%%%%%%%%%%%%%%%%%%%%%%%%%%%%%%%%%%%%

To further  illustrate these results, we show in Fig.~\ref{wfun} a typical spatial image of an eigenfunction for disorder $W=10$. The area of each circle in this figure is proportional to the  absolute value $|\psi|$ of the wave function amplitude on the corresponding site. One can indeed observe that the fluctuations become much stronger when one moves from the root towards the boundary.

\section{Summary}

In this paper, we have explored the evolution of wave-function statistics on a Cayley tree from the central site (``root'') to the boundary (``leaves'').
We have shown that the eigenfunction moments  (\ref{Pq})  exhibit a multifractal scaling (\ref{Pq-scaling})   with the number of sites $N$ at $N\to\infty$. The multifractality spectrum $\tau_q$ depends on the strength of disorder and on the parameter $s$ characterizing the position on the lattice. Specifically, $s= r/R$, where $r$ is the distance from the observation point to the root, and $R$ is the ``radius'' of the lattice. Using analytical and numerical approaches, we have studied the evolution of the spectrum with increasing disorder, from delocalized to the localized phase. 

The analytical results have been obtained for the $n$-orbital model with $n \gg 1$ that is mapped onto a supersymmetric $\sigma$ model. We have derived recurrence relations that determine the distribution function of wave-function amplitudes at an arbitrary position on the lattice and have analyzed the scaling of the corresponding moments. The key findings are as follows:

\begin{enumerate}

\item We have determined the fractal exponents $\tau_q$ as functions of $q$, the position $s$, and the coupling constant $g$ (that carries the information about the disorder strength).  There are four domains in the plane spanned by $q$ and $g$, with distinct analytical formulas for $\tau_q$, see Fig.~ \ref{pd} and Sec.~\ref{sec:phase-diagram}. 

\item The exponents $\tau_q$ depend linearly on the position $s$ on the tree.

\item The statistics on the boundary of the tree, $s=1$, shows a characteristic property of  localization for any $g$. Specifically, we find $\tau_q = 0$ for all $q$ above a certain value (that was denoted $Q$ or $Q_*$ depending on the value of $g$, see Fig.~ \ref{pd}). This implies that, while $g > g_c$ is the delocalized phase on an infinite Bethe lattice (and on RRG), eigenfunctions on a finite Cayley tree  are localized at $g>g_c$ near boundary sites. 

\item Despite the localized character of high moments at $g < g_c$ (any $s$) as well as at $s=1$ (any $g$), the multifractal behavior of low moments persists also in these regimes.

\end{enumerate}

These results have been supported by exact diagonalization of the conventional ($n=1$) Anderson tight-binding model. The analytical and exact-diagonalzation results show a good agreement in most of salient features, see \ref{rdep}, \ref{qdep5}, \ref{qdep8}, \ref{qdep14}, \ref{qdep18}, and   \ref{dt}. This agreement is not  trivial, since (i) the analytical and numerical data correspond to somewhat different models ($n\gg 1$ and $n=1$ Anderson models, respectively) and (ii) numerical data are subjected to finite-size effects. The numerics confirms, in particular, the localized character of high moments  and the fractal character of low moments at the boundary, $s=1$, for any disorder. The former result (the found behavior of high moments at $s=1$) is consistent with the localization of eigenstates on canopy graphs that has been rigorously proven in Ref.~\onlinecite{aizenman2006canopy}. 

Two additional comments are in order here:

\begin{enumerate}

\item Our work demonstrates once more a crucial difference between the physics of eigenstates on finite trees, on one hand, and on tree-like graphs without boundary (RRG and SRM models), on the other hand. This difference was not always appreciated in the recent literature\cite{altshuler2016nonergodic,altshuler2016multifractal}. 

\item The linear dependence of multifractal exponents $\tau_q$ on the parameter $s=r/R$ on the Cayley tree bears analogy with the linear dependence of multifractal exponents at the Anderson transition point in a $d$-dimensional system ($d<\infty$) on the parameter $\beta = \ln \rho / \ln R$, where $R$ is the system size, and $\rho$ the distance to the boundary \cite{subramaniam2006surface}. Within this analogy, $1-s = \rho/R$ corresponds to $\beta$, with zero value of these parameters corresponding to the surface multifractality and the value unity to the bulk multifractality. In both cases, the states become ``more fractal'' (i.e. closer to localized) when one moves from the  central part of the system towards the boundary. The fact that $\ln \rho / \ln R$ for finite dimensionality translates into $\rho/R$ for Cayley tree is related to the fact that the volume is a power-law function of the linear size in the first case and an exponential function in the second case. 

\end{enumerate}

Before closing the paper, we briefly discuss prospective directions for future research.
%\begin{enumerate}

%\item
First,  it will be interesting to study manifestations of multifractality of eigenfunctions in spatial and dynamical correlations. In particular, the analysis of correlations should provide a more precise quantitative characterization for the spatial structure of eigenfunctions illustrated in Fig. \ref{wfun}.

%\item 
A second important direction is to extend our analytical study of multifractality on a finite Cayley tree onto the $n=1$ Anderson model. This should show whether all non-analiticity points and lines in Figs.~\ref{figlambda} and \ref{pd} retain their character for $n=1$ or some of them become crossovers. In fact, the analysis in Ref.~\onlinecite{aizenman2006canopy} suggests that the moments in the $n=1$ Anderson model may show non-ergodicity at root at any disorder, at variance with the ergodic-at-root domain found at $g > g_e$ in the $\sigma$ model ($n \gg 1$ limit of the Anderson model). 

%\item 
Third, as was mentioned in Sec.~\ref{SectionIntro}, Anderson models on tree-like graphs attract much attention currently in view of their similarity to Fock-space representation of disordered interacting many-body problems. In this context, tree-like graphs without a boundary (like RRG and SRM models) are of particular relevance. It remains to be seen whether finite Cayley trees, and in particular the eigenstate multifractality explored in the present work, may find some applications in the context of many-body physics. 

Fourth, it is worth pointing out that, in addition to finite Cayley trees studied in this work and in Ref.~\onlinecite{tikhonov2016fractality}, there exists a number of models that show phases (rather than isolated points as in the case of conventional Anderson transitions) with multifractal eigenstates. These include, in particular, power-law random banded matrices\cite{evers08}, small-world networks \cite{giraud2005quantum,quintanilla2007electron}, Rosenzweig-Porter model \cite{kravtsov2015random,facoetti2016non,von2017non}, and random Levy matrices \cite{monthus2016localization}. It would be interesting to see whether there are deeper connections between all these models (or, at least, some of them).

Finally, let us note that we have derived statistical properties of wave functions on a Cayley tree from the analysis of Eq. (\ref{PiRecursionSimple}), which describes a linear integral recursion with the kernel determined by the solution of a Fisher-KPP-like equation, Eq. (\ref{PsiRecursionSimple}). In view of a wide area of applicability of equations of Fisher-KPP type, it would be interesting to understand the role of equations analogous to Eq. (\ref{PiRecursionSimple}) in a broader context of front propagation problems. In particular, a similar equation (although with very different initial condition) arises in the analysis of distribution of a distance between the extreme points in the problem of branching Brownian motion \cite{brunet2011branching}.
%\end{enumerate}

\section{Acknowledgements}

We thank M. Aizenman, Y. Fyodorov and S. Warzel for discussions. A.D.M. acknowledges the hospitality of the Weizmann Institute of Science within the Weston Visiting Professorship. This work was supported by Russian Science Foundation under Grant No. 14-42-00044. K.T. was also partially supported by the RF Presidential Grant No. NSh-10129.2016.2

\bibliography{bethe}

\end{document}